\documentclass[journal=jctcce,manuscript=article,layout=traditional]{achemso}

\usepackage{amsmath,amssymb,amsthm}
\usepackage{array}
\usepackage{booktabs}
\usepackage{longtable}
\usepackage{microtype}
\usepackage{tikz}
\usetikzlibrary{arrows.meta,calc,positioning}

\providecommand{\href}[2]{#2}
\providecommand{\url}[1]{#1}
\providecommand{\keywords}[1]{\par\noindent\textbf{Keywords: }#1\par}
\makeatletter
\@ifundefined{tocentry}{%
  \newenvironment{tocentry}{\section*{Graphical Table of Contents}}{}%
}{}
\makeatother

\newcommand{\SONIC}{SONIC}
\newcommand{\SMITH}{\textsc{SMITH}}
\newcommand{\GIC}{GIC}
\newcommand{\Bmat}{\mathbf{B}}
\newcommand{\bmat}{\mathbf{b}}
\newcommand{\Cmat}{\mathbf{C}}

\newcommand{\Gmat}{\mathbf{G}}
\newcommand{\x}{\mathbf{x}}
\newcommand{\q}{\mathbf{q}}
\newcommand{\y}{\mathbf{y}}

\theoremstyle{definition}
\newtheorem{definition}{Definition}

\title{\SONIC: Symmetry-Oriented Non-redundant Internal Coordinates}
\author{Vincenzo Barone}
\affiliation{INSTM, National Interuniversity Consortium for Materials Science and Technology, via G. Giusti 9, 50121 Firenze, Italy}
\email{vincebarone52@gmail.com}

\begin{document}
\maketitle

\begin{abstract}
Internal coordinates are the natural language of molecular structure, yet their
automatic construction remains uneven across the use cases required by modern
spectroscopy and structure refinement.  Primitive redundant internals are robust
for geometry optimization but do not define a unique active space.  Global
non-redundant coordinates obtained from the singular-value or eigenvalue
decomposition of a full primitive Wilson matrix remove the redundancy, but the
resulting variables are often delocalized and sensitive to arbitrary weights
assigned to stretches, bends, torsions and non-covalent contacts.  Natural
internal coordinates provide the best chemical interpretation, especially for
force-field and vibrational work, but existing constructions are usually tied to
particular bonding patterns and do not provide a general, reproducible treatment
of fused rings, high coordination, point-group symmetry and weakly bound
fragments.

We present \SONIC\ (Symmetry-Oriented Non-redundant Internal Coordinates) as a general
construction of non-redundant, symmetry-adapted generalized internal
coordinates.  The \SMITH\ (Symmetry-Mapped Internal-coordinate Template
Handler) implementation starts
from Cartesian geometry and one of four provider-neutral input profiles: a
complete frozen molecular state, user-supplied topology, user-supplied
redundant primitives, or a standalone perception kernel that constructs
ordinary topology, molecular point-group operations and primitive coordinates.
It validates and classifies
the ordinary primitive rows, augments them with chemically typed special or
composite families, protects coordinates for rings and fragments, and reduces the coordinate space using analytic Wilson
\(\Bmat\)-row rank tests.  A local-pseudosymmetry layer first orders
center-, ring- and bond-domain coordinates with the locally totally symmetric
combination first; the exact molecular point-group projector is then applied
inside homogeneous coordinate blocks.  The result is a deterministic constructive protocol and a
frozen coordinate contract, rather than an arbitrary non-redundant basis.  The
contract can also incorporate optional user-defined protected semantic
coordinates and derived observables.  The published SMITH boundary ends with
the coordinate definitions, their human-readable decomposition, symmetry and
rank diagnostics, and analytic Wilson \(\Bmat\) rows.  Optimization policies,
finite internal-to-Cartesian realization, force-field models and higher-order
Hessian transformations are consumers of this contract, not parts of its
construction.  The method
therefore provides a constructive counterpart to natural internal coordinates:
it preserves their locality and interpretability while extending them to the
topological and symmetry cases that usually require manual intervention.
In MATRIX, molecular symmetry and ordinary PICs are authoritative ORACLE input
and are never reperceived by SMITH.  To remain independently installable, the
standalone distribution embeds a revision-pinned subset of the same ORACLE
perception kernel.  The duplication is therefore a packaging boundary, not a
second symmetry method, and its provenance is recorded with the SONIC contract.
The numerical evidence reported here validates the construction invariants,
sparse first-derivative machinery and external serialization path.  Gaussian 16
is used because its unusually general GIC language can evaluate the generated
SONIC expressions directly, providing an independent parser and solver test
without molecule-specific transcription.  It is not used as a competing
coordinate generator.  Its export limitations are explicit: native
out-of-plane and special composite coordinates remain in the SONIC contract
but must be translated to supported improper-dihedral or component expressions,
or omitted from the commercial input when no exact GIC representation exists.
Standalone probes include formic-acid--water, water-dimer and benzene--water
complexes with six protected intermolecular degrees of freedom, together with
an $\eta^3$ allyl--palladium complex in which a supplied ligand center is
retained as a protected metal--center coordinate.  We also discuss the
intended scope of the method and its current limitations.
\end{abstract}

\begin{tocentry}
\centering
\includegraphics[width=0.9\linewidth]{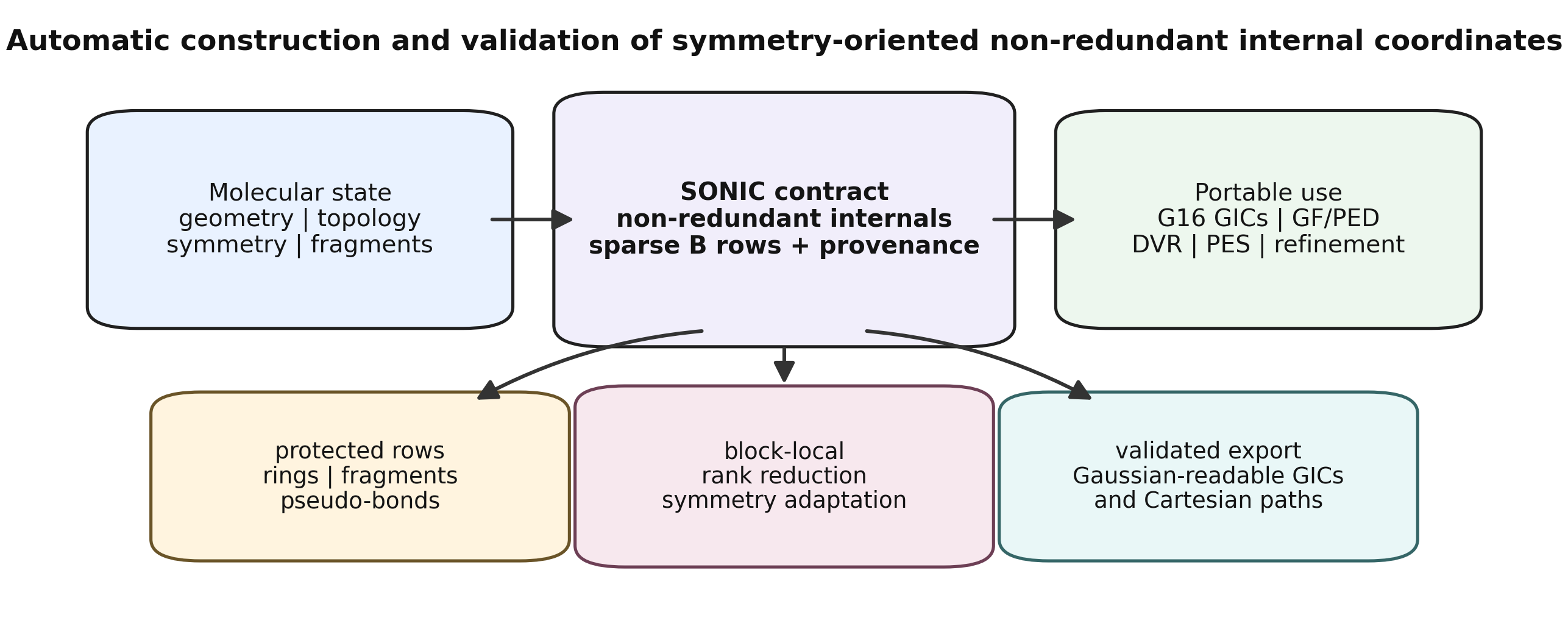}
\end{tocentry}

\section{Introduction}

The Wilson internal-coordinate formalism gives a compact relation between
Cartesian displacements \(\Delta \x\) and internal-coordinate displacements
\(\Delta \q\),
\begin{equation}
  \Delta \q = \Bmat \Delta \x ,
\end{equation}
and remains the standard framework for vibrational analysis, potential-energy
distributions and internal-coordinate force constants.\cite{Wilson1955}  The
formalism itself does not solve the constructive problem: one must still choose
which coordinates define the molecular model.  That choice is benign in small
hand-built Z matrices, but it becomes a central scientific issue when the same
coordinate model is expected to support symmetry-constrained optimization,
semi-experimental structural refinement, ring puckering, weak complexes and
metal or high-coordination environments.

\SONIC\ is designed to make this choice explicit and reproducible.  Table~\ref{tab:framework-roles}
defines the permanent boundary used below.  The essential separation is that
\SMITH\ is the standalone tool that constructs \SONIC\ contracts, while
\SONIC\ is the coordinate family: the non-redundant generalized
internal-coordinate basis, analytic
Wilson matrix and symmetry labels.  Applications choose how those coordinates
are displaced, optimized, fitted or differentiated beyond first order.

\begin{table}[htbp]
\centering
\caption{Permanent boundary of the standalone SMITH/SONIC work.}
\label{tab:framework-roles}
\begin{tabular}{@{}p{0.18\linewidth}p{0.72\linewidth}@{}}
\toprule
Name & Role \\
\midrule
Input provider & Supplies Cartesian geometry and, optionally, topology,
redundant primitives, symmetry operations, fragments, interaction centers and
continuous chemical descriptors.  For plain Cartesian input, standalone SMITH
embeds a revision-pinned perception subset for ordinary topology, point-group
operations, atom permutations and primitive coordinates. \\
\SMITH{}--\SONIC & Coordinate-construction tool and its coordinate
contract.  \SMITH\ accepts a complete frozen state, a supplied topology, a
supplied redundant primitive set, or plain Cartesian input through its bundled
perception frontend.  It performs SONIC construction, special/composite coordinate
generation, symmetry adaptation, analytic \(\Bmat\) evaluation, human-readable
reporting, Gaussian-16-compatible export and provenance. \\
Application layer & Consumes the frozen contract.  Internal-to-Cartesian
realization, optimization, scanning, force-field construction, property
models, Hessian transport and higher derivatives are deliberately outside
SMITH. \\
\bottomrule
\end{tabular}
\end{table}

Within the integrated MATRIX implementation these generic roles have one
explicit mapping, stated here once to avoid coupling the standalone method to
the suite architecture. ORACLE supplies symmetry, atom permutations, redundant
primitives and their Wilson matrix; SMITH consumes that frozen state and
constructs SONIC without reperception; LINK owns finite internal-to-Cartesian
realization for optimization, scans and externally proposed points; ARCHITECT
constructs $\Bmat'$ on demand for nonstationary Hessian transformations;
MORPHEUS delegates any coordinate realization to LINK; and SENTINEL exchanges
SONIC points with LINK without consuming $\Bmat$.

The standalone executable preserves this scientific boundary by copying only
the required perception implementation into its release snapshot.  It can
therefore infer symmetry from an XYZ file without installing ORACLE, but the
algorithm, thresholds and schema remain those of the pinned ORACLE revision.
If a complete frozen state or an explicit symmetry section is supplied, that
record is authoritative and the embedded kernel does not replace it.

\begin{figure}[htbp]
\centering
\resizebox{0.99\linewidth}{!}{%
\begin{tikzpicture}[
  node distance=0.46cm and 0.56cm,
  every node/.style={font=\normalsize},
  stage/.style={
    draw=black!65,
    line width=0.45pt,
    rounded corners=2.2pt,
    align=center,
    text width=0.155\linewidth,
    minimum height=1.16cm,
    inner sep=4.8pt,
    fill=#1
  },
  contract/.style={
    draw=black!80,
    line width=0.65pt,
    rounded corners=2.4pt,
    align=center,
    text width=0.18\linewidth,
    minimum height=1.26cm,
    inner sep=5.0pt,
    fill=black!6
  },
  app/.style={
    draw=black!50,
    line width=0.35pt,
    rounded corners=2pt,
    align=center,
    text width=0.14\linewidth,
    minimum height=0.74cm,
    inner sep=3.4pt,
    fill=black!2,
    font=\small
  },
  arrow/.style={-{Latex[length=2.2mm]}, line width=0.45pt, draw=black!70}
]
\node[stage=blue!7] (state) {Frozen state\\or embedded\\perception};
\node[stage=green!8, right=of state] (families) {Primitive/B\\source + SMITH\\family labels};
\node[stage=orange!10, right=of families] (semantic) {Protected\\semantic, ring\\and fragment rows};
\node[stage=purple!8, right=of semantic] (reduction) {Block-local\\rank reduction\\and symmetry};
\node[contract, right=of reduction] (contract) {Frozen\\coordinate\\contract};

\node[app, below=1.00cm of contract, xshift=-0.32\linewidth] (opt) {Optimization};
\node[app, right=0.14cm of opt] (gf) {GF/PED};
\node[app, right=0.14cm of gf] (dvr) {DVR};
\node[app, right=0.14cm of dvr] (pes) {PES fitting};
\node[app, right=0.14cm of pes] (refine) {Refinement};

\draw[arrow] (state) -- (families);
\draw[arrow] (families) -- (semantic);
\draw[arrow] (semantic) -- (reduction);
\draw[arrow] (reduction) -- (contract);
\coordinate (bus) at ($(contract.south)+(0,-0.42cm)$);
\draw[arrow] (contract.south) -- (bus);
\draw[draw=black!55, line width=0.35pt] (opt.north |- bus) -- (refine.north |- bus);
\foreach \target in {opt,gf,dvr,pes,refine}
  \draw[arrow] (\target.north |- bus) -- (\target.north);
\end{tikzpicture}
}
\caption{Constructive workflow implemented by \SMITH\ for \SONIC\ coordinates,
starting from Cartesian geometry and an optional topology or redundant
primitive/B contract.  Standalone SMITH can construct the minimal input state
or accept richer provider-generated sections without changing its own boundary.
The model-building stage
ends at the frozen coordinate contract; downstream modules reuse that object
but their algorithms are not part of SMITH.}
\label{fig:workflow}
\end{figure}
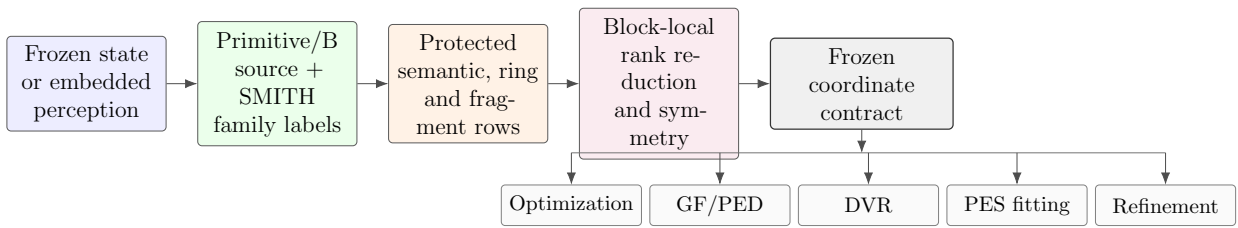

The novelty is not a new coordinate formula, but a deterministic constructive
protocol.  A useful generator must choose a non-redundant basis while satisfying
four requirements at once: correct vibrational rank, chemical locality,
symmetry adaptation without cross-family mixing, and protection of special
coordinates such as ring puckering components, ring-center distances, fragment
translations and fragment orientations.  The result is not merely a coordinate
list but a persistent object that downstream modules can audit, reuse and
compare against independent active-space constructions.

Equally important, \SONIC\ is not a new coordinate language.  Modern GIC
implementations already provide expressive syntax for nonlinear and mixed
coordinate expressions.  The novelty here is the preceding step: a
deterministic generator that converts a frozen molecular state into a
rank-complete, symmetry-consistent coordinate basis that can be serialized as
Gaussian-readable GICs or consumed directly by downstream modules.  The
standalone command-line tool is named \SMITH\ (Symmetry-Mapped
Internal-coordinate Template Handler).  Symmetry is used in the classical molecular sense: point-group
operations map coordinate rows into the same space with the corresponding atom
permutation and vector or pseudovector signs.  Orthogonality enters through
normalized Wilson-row rank selection and orthonormal SALC coefficient vectors
inside homogeneous blocks, not through global orthogonality in the kinetic
\(\Gmat\) metric.

\section{Relationship to Existing Coordinate Models}

Table~\ref{tab:context} summarizes the position of \SONIC\ relative to the main
coordinate strategies used in quantum chemistry and molecular spectroscopy.
The comparison is deliberately constructive.  \SONIC\ is not a replacement for
the Wilson formalism, natural internal coordinates or translation-rotation
coordinates; it is a general layer that decides how such ingredients are
assembled into one frozen, reusable coordinate object.

\begin{table}[htbp]
\centering
\scriptsize
\setlength{\tabcolsep}{3pt}
\renewcommand{\arraystretch}{0.92}
\caption{How \SONIC\ is positioned relative to established coordinate strategies.}
\label{tab:context}
\begin{tabular}{p{0.19\linewidth}p{0.31\linewidth}p{0.39\linewidth}}
\toprule
Strategy & Strength & Limitation addressed by \SONIC \\
\midrule
Z matrices & Compact, local and interpretable for small molecules. & Tree
dependence, singularities, and manual choices make them fragile for rings,
symmetry and automated workflows.\cite{BakerHehre1991ZMatrix} \\
\midrule
Redundant internal coordinates & Robust for optimization and easy to generate
from a graph.\cite{PulayFogarasi1992,Peng1996Redundant,BakkenHelgaker2002} &
They do not define a unique active basis for vibrational analysis, symmetry
projection or least-squares refinement. \\
\midrule
Delocalized internal coordinates & Automatic non-redundant bases can be
obtained from global rank decompositions.\cite{Baker1996Delocalized} & The
coordinates can mix stretches, bends, torsions and weak contacts; weights used
to control this mixing are method choices rather than chemical observables. \\
\midrule
Natural internal coordinates & Highly interpretable symmetry and force-field
variables for well understood local motifs.\cite{Berces1997NaturalMetal,VonArnimAhlrichs1999GNIC,SwartBickelhaupt2006StrongWeak}
& General construction is difficult for fused or bridged rings, high
coordination, fragments, interaction centers and arbitrary point groups. \\
\midrule
TRIC/geomeTRIC coordinates & Explicitly represents collective fragment
translations and rotations, with the rotations described by an exponential map
of quaternions.\cite{WangSong2016GeomeTRIC} & \SONIC\ adopts this
exponential-map parametrization, but not the complete TRIC construction: it
defines the rigid-body directions relative to a reference fragment and retains
them as a protected, typed and symmetry-labelled source family; alternatively,
the intermolecular model may be constructed from typed pseudo-bonds and
pseudo-cycles. \\
\midrule
Gaussian GIC syntax & Very general symbolic coordinate expressions, including
nontraditional geometric parameters, sparse derivative rows and automatic
symbolic differentiation, can be evaluated by an external optimizer; the recent
Gaussian implementation is the most direct general-purpose
comparison.\cite{Marenich2025GIC} & Standard internals can be generated from
connectivity and user-defined GIC expressions can be parsed and differentiated,
but the chemically specialized examples remain coordinates that are explicitly
specified in input or activated through dedicated options.  The syntax does not
define a complete natural-coordinate generator with protected rank reduction,
homogeneous point-group projectors and reusable symmetry-labelled blocks for
refinement or force-field analysis. \\
\midrule
\SONIC & Builds a rank-complete, type-labelled, symmetry-adapted coordinate basis
from a validated molecular state. & Weak complexes can be represented either
by fragment/center coordinates or by typed pseudo-bonds and
pseudo-cycles; the remaining development frontier is the breadth of the
pseudo-cycle validation corpus. \\
\bottomrule
\end{tabular}
\end{table}

This positioning is important for the narrative of the method.  Primitive
redundant internals and delocalized internals are excellent tools for geometry
optimization, where the main criterion is stable movement on a potential-energy
surface.  A semi-experimental fit or a Wilson GF analysis imposes stricter
requirements: the fitted or analyzed variables must be identifiable, symmetry
consistent and auditable as the same objects across independent modules.
Natural internal coordinates meet these requirements when the local chemistry is
known in advance.  B{\'e}rces extended their use to transition-metal skeletal
degrees of freedom,\cite{Berces1997NaturalMetal} von Arnim and Ahlrichs
generalized the automatic construction while still identifying multiply fused
ring systems as a hard case,\cite{VonArnimAhlrichs1999GNIC} and Swart and
Bickelhaupt emphasized the need to distinguish strong and weak coordinates in
optimization.\cite{SwartBickelhaupt2006StrongWeak}  \SONIC\ turns this lineage
into a frozen construction for spectroscopy and refinement by combining
chemically typed primitive generation, block-local rank reduction,
special-coordinate protection and point-group projectors.

The recent Gaussian GIC paper is therefore the closest comparison and also
clarifies the remaining gap.  Its Supporting Information demonstrates the
power of the GIC language with nonlinear water coordinates, formamide
proton-transfer sum and difference variables, methane and ammonia
\(XY_3\)-type symmetrized combinations, polar and elliptic Li\(_2\)--Ar scan
variables, fragment rotor coordinates for benzene dimers and water clusters,
and quaternion fragment coordinates for ferrocene.\cite{Marenich2025GIC}
These examples clarify the distinction between an expressive coordinate
language and an automatic generator.  Gaussian can evaluate and differentiate
such expressions once they have been supplied or selected.  \SONIC\ instead aims
to generate the corresponding objects whenever they follow from topology, local
equivalence, point-group symmetry, fragments, ring centers, bond centers or
semantic coordinate requests.

This distinction is central to the positioning of the method.  Gaussian GIC is
an expressive coordinate-description language: a user can write arbitrary
primitive, nonlinear, redundant or mixed expressions and let Gaussian evaluate
them.  \SONIC\ addresses a different problem.  It constructs the coordinate model
before any optimizer, force-field fit or refinement sees it, and records why
each special or ordinary coordinate was accepted.  The Gaussian syntax is
therefore a serialization target for \SONIC, not the theoretical object being
introduced.

In practical terms, the user should not have to write pages of molecule-specific
GIC definitions to obtain chemically natural variables such as local
\(XY_n\)-type combinations, proton-transfer sums and differences, fragment
translations or ring-puckering components.  When those variables are consequences
of the frozen molecular state, \SONIC\ can generate and protect them
automatically; when they are requested explicitly, the request is recorded as a
semantic coordinate rather than as an opaque handwritten expression.

\section{Molecular State and Atom Equivalence}

\SONIC\ does not rediscover the molecule at every downstream step.  Its formal
input boundary is provider-neutral: Cartesian geometry plus either a frozen
topology or an ordered redundant primitive set with the information needed to
reconstruct its Wilson B rows.  A complete molecular state may additionally
contain rings, point-group operations, fragments, interaction centers and
continuous chemical-perception descriptors.  If those data are already
present in the \texttt{xyzin} file, \SONIC\ consumes them as part of the frozen
state.  The provider may be the bundled perception frontend, another program or a
hand-audited input; its identity is not part of the SONIC theory.  The bundled
frontend includes the revision-pinned symmetry, topology and ordinary-PIC
subset required for standalone construction.  A rich state
can define the molecular graph and cycle basis, point-group operations and atom
permutations, atom equivalence, effective atomic numbers, continuous charge,
covalency, delocalization and strain descriptors, bond-order components and
synthon indices.  SMITH consumes these records when present.  The complete
state may also supply ordinary stretch, bend, linear-bend, dihedral and
out-of-plane source definitions and their reference B matrix.  SMITH adds only
the fragment, pseudo-bond, ring-puckering, interaction-center or protected
semantic rows required by the requested SONIC chart.
The output is a frozen \(\#\)GIC definition: consumed and augmented primitive coordinate records,
final generalized coordinates, coefficients, family labels, irreducible
representations, total-symmetric subsets and the metadata required to
reconstruct analytic \(\Bmat\) rows.  This is the end of the SMITH contract.

For standalone use the required boundary can be supplied in a single
\SMITH\ extended-XYZ input file with schema \texttt{matrix.smith.input.v1}.  The file begins as a standard XYZ block: atom count, comment line and Cartesian
coordinates.  Optional SMITH directives follow after the atom list, so the same
file can still be opened by ordinary XYZ viewers while carrying construction
options such as symmetrization, \texttt{SYCART} export, fragment treatment,
X--H policies and an optional reduced working subgroup.  Four input profiles
are accepted.  A complete frozen state is consumed unchanged; a supplied
\texttt{TOPOLOGY} section is used to generate the ordinary redundant
primitives; a supplied \texttt{PRIMITIVES} section is validated against the
Cartesian geometry and used directly; and a plain XYZ invokes a bundled
perception frontend that constructs topology, a complete molecular point-group
action with atom permutations, and ordinary primitives.  If
topology and primitives are both supplied, their bond rows must agree.  The
output records the selected profile as \texttt{FROZEN\_STATE},
\texttt{STANDALONE\_TOPOLOGY}, \texttt{STANDALONE\_PRIMITIVES} or
\texttt{STANDALONE\_MINIMAL}.  The minimal route is sufficient for ordinary
molecules and the supplied examples; the profile name is retained for schema
compatibility even though molecular symmetry is included.  Advanced continuous
descriptors, interaction centers, user-selected quasi-symmetry policies or an
externally audited symmetry state can instead be supplied through the
complete-state interface.  Thus SMITH is operationally standalone while
preserving a precise boundary between perception and coordinate construction:
the embedded subset provides the input state, and SMITH proper begins only
after that state has been frozen.  The four complete input examples in the
Supporting Information are executable specifications of this boundary.

The richer optional descriptors follow continuous molecular perception
introduced in Proxima,\cite{Lazzari2020Proxima} and molecular/spectral synthons developed for accurate structural
templating.\cite{Gambi2019Synthons,Lazzari2024SynthonPCS,Lazzari2025PCS141}
Initial local equivalence is therefore not based only on element symbols.  The
input state may supply continuous atom-perception and synthon
descriptors, and \SONIC\ uses them to make reproducible local choices.
The present construction uses effective atomic numbers
that combine nuclear charge with local coordination, distance-derived
bond-order information, angular rigidity and delocalization-like terms.  These
continuous descriptors order atoms and ligands in cases where a pure graph
label would be ambiguous.  The topology layer also exposes synthon
descriptors--charge, covalency, delocalization, strain, effective nuclear
charge and incident $\sigma/\pi/\pi\pi$ indices--which provide a natural route
to more refined equivalence
classes.  In the present contract, this continuous perception enters ring
orientation, ligand grouping, local stretch equivalence, high-coordination
templates and the one-dihedral exocyclic torsion rule.  In particular, ring
puckering diagnostics report the first- and second-$\pi$ indices of every
endocyclic central bond alongside its topological order and flexibility factor.
Triple-bond character is therefore not collapsed into the same aggregate
$\max(\mathrm{BO}-1,0)$ used historically for all multiple bonds.
The reproducibility statements below are conditional on these descriptors being
part of the frozen input state.  In other words, \SONIC\ does not assume an
unstated human perception step: the descriptor values, perception version,
topology and symmetry operations are stored as contract provenance, and a
change in any of them is a different molecular-state stratum.

\section{Coordinate Functions and Downstream Models}

The frozen \SONIC\ basis is the first-order coordinate frame.  Many downstream
models, however, are better written in functions of those coordinates than in
the raw coordinates themselves.  This is especially true for force fields and
PES fitting, where the physically smooth variable may be an inverse distance,
an exponential contact coordinate, a trigonometric torsional term or a
ring-puckering descriptor.  We therefore distinguish base coordinates
\(\q\), which define rank and symmetry, from model coordinates
\(\y=f(\q)\), which are evaluated by downstream modules.

For any differentiable coordinate function \(y_a=f_a(\q)\), the Wilson row used
by a fit or force-field model follows from the chain rule,
\begin{equation}
  \label{eq:chain-rule}
  \frac{\partial y_a}{\partial \x}
  =
  \sum_i
  \frac{\partial f_a}{\partial q_i}
  \frac{\partial q_i}{\partial \x}.
\end{equation}
Thus the same frozen \(\Bmat\) service can support nonlinear coordinate
functions without changing topology, rank selection or symmetry labels.
Examples include \(r^{-1}\), \(\exp(-\alpha r)\) and damped contact functions
for distances; \(\sin\theta\), \(\cos\theta\), \(\sin n\phi\) and
\(\cos n\phi\) for bends and torsions; and Cremer--Pople-like \(q\) and
\(\phi\) descriptors derived from linear \(RPck\) components.  The same layer
also covers the practical variables used in the Gaussian GIC examples:
bond-sum and bond-difference coordinates for proton transfer, symmetrized
\(XY_n\) local combinations, polar or elliptic functions of intermolecular
Cartesian components, and quaternion or exponential-map functions of fragment
orientation.\cite{Marenich2025GIC}  In a GIC input language these are written
as explicit expressions or requested through a specific option; in \SONIC\ they
are downstream functions of generated base coordinates or protected fragment
coordinates, with the chain rule preserving the connection to the frozen
Wilson rows.  Recent work by
Lema-Saavedra and Fern{\'a}ndez-Ramos shows why this separation is useful for
rings: Cremer--Pople amplitudes provide an efficient conformational search
space, while chemically viable 3D geometries require ring closure, angular
redistribution, treatment of rigid or multiple bonds and invariant local
references for substituents.\cite{LemaSaavedraFernandezRamos2026RingCP}  The
active Gaussian-readable coordinate basis should remain linear and rank
complete, while such nonlinear functions become named observables or model
variables in GF/PED, DVR, PES-fitting or refinement workflows.

Conformational reconstruction from puckering variables is therefore not external
to this workflow.  In the DVR layer, a one- or multidimensional grid in
\(RPck\)-, \(Q\)- or \(\phi\)-like variables must be associated with physically
closed ring geometries before electronic-structure points, metric terms or
wavefunctions can be interpreted.  The reconstruction layer of
Ref.~\citenum{LemaSaavedraFernandezRamos2026RingCP} provides precisely this
coordinate-to-geometry step: closure constraints, angular redistribution,
rigid-bond treatment and substituent reference frames belong above the frozen
\SONIC\ basis, but they are natural consumers of it.  A DVR treatment can then
recompute the Wilson metric on the reconstructed structures, or use an explicitly
validated reduced metric, without introducing a second private ring-coordinate
definition.

This distinction also prevents a common ambiguity in the word ``GIC.''  A
general GIC expression can be an optimizer coordinate.  A \SONIC\ coordinate
definition is reusable: it specifies the base coordinate
space, the analytic derivatives and the symmetry transformation rules from
which later functional variables can be built.  Implementing these coordinate
functions systematically is part of the next development layer, but it does not
require a different coordinate-generation theory.

\SONIC\ therefore exposes a small semantic coordinate layer rather than a second GIC
language.  \texttt{PROTECT} requests name base objects such as distances,
torsions, ring puckering, butterfly coordinates, fragment translations or
center distances.  Independent requests enter rank and symmetry selection with
the same priority as automatically generated protected coordinates.
\texttt{OBSERVABLE} requests define functions such as inverse distances,
trigonometric terms, linear combinations, Cremer--Pople variables or
proton-transfer coordinates; they are evaluated later as \(y=f(\q)\) by
Eq.~\eqref{eq:chain-rule}.  Duplicate user rows are rejected,
row-equivalent user rows substitute the automatic primitive with provenance
recorded, and partial symmetry orbits are diagnosed before local SALC fallback.

\section{The Frozen Coordinate Contract}

The central output of \SONIC\ is a persistent coordinate object, not a transient
list of internal coordinates.  This distinction is important because the same
coordinate object must be used by independent modules, with no private
reconstruction of topology, symmetry or rank choices.

\begin{definition}
\label{def:frozen-contract}
A frozen coordinate contract is the tuple
\[
  \mathcal{C} = (\mathcal{P},\mathcal{G},\Bmat,\Sigma,\mathcal{M}).
\]
Here \(\mathcal{P}_{0}=(p_1,\ldots,p_{m_0})\) is the ordered redundant
primitive/B source supplied at the input boundary and \(\mathcal{P}\supseteq\mathcal{P}_{0}\)
is the typed source space after any SMITH special/composite augmentation;
\(\mathcal{G}=(\mathbf{g},\Cmat_{\mathcal{G}})\) is the accepted
non-redundant generalized-coordinate basis together with its coefficient
matrix; \(\Bmat\) contains the analytic Wilson rows of the accepted basis;
\(\Sigma\) contains point-group transformation matrices,
irreducible-representation labels and SALC coefficients; and \(\mathcal{M}\)
contains the frozen molecular state, source-family labels, protection flags,
family priority order, numerical tolerances, semantic grammar and contract
schema versions, diagnostics and provenance.
Equivalently, with \(\mathbf{p}=(p_1,\ldots,p_m)^T\),
\[
  \mathbf{g}=\Cmat_{\mathcal{G}}\mathbf{p},
  \qquad
  g_a=\sum_{j=1}^{m} c_{aj}p_j,
  \qquad
  \Bmat=\frac{\partial\mathbf{g}}{\partial\x}
       =\Cmat_{\mathcal{G}}\Bmat(\mathcal{P}).
\]
\end{definition}

The tuple is the object consumed by GF/PED, DVR, PES-fitting and refinement
workflows.
It is also the object that makes the method auditable: a downstream result can
be traced to the primitive candidates that were generated, the rows that were
accepted or rejected, the rank tolerance that was used and the symmetry block in
which each final coordinate was projected.

Let \(X=\mathbb{R}^{3N}\) be the Cartesian displacement space and let \(T\)
denote the rigid-body tangent space generated by overall translations and
rotations.  For a nonlinear molecule, \(\dim T=6\); for a linear molecule,
\(\dim T=5\).  Scalar intramolecular coordinates have first-order rows in the
annihilator
\[
  X_{\mathrm{int}}^*
  =
  \{\ell\in X^*:\ell(t)=0\ \text{for all}\ t\in T\},
\]
whose target dimension is the vibrational rank \(r=3N-\dim T\).  Vector-like
fragment translations and orientations are stored as relative or body-fixed
generalized coordinates and are audited against the same external-mode-free
active space.  Each primitive candidate \(p_i\in\mathcal{P}\) contributes an
analytic row \(\bmat_i=\partial p_i/\partial\x\), and
\[
  \Bmat(\mathcal{P}) =
  \begin{bmatrix}
    \bmat_1 \\
    \vdots \\
    \bmat_m
  \end{bmatrix},
  \qquad
  \mathcal{R}(\mathcal{P}) =
  \operatorname{span}\{\bmat_1,\ldots,\bmat_m\}.
\]
The constructive goal is to choose an ordered subset
\(\mathcal{A}\subseteq\mathcal{P}\) such that
\[
  \operatorname{span}\Bmat(\mathcal{A})=\mathcal{R}(\mathcal{P}),
  \qquad
  |\mathcal{A}|=\operatorname{rank}\Bmat(\mathcal{P})=r ,
\]
when the primitive candidate set spans the full internal row space.  If this
condition fails, the build is not silently repaired; the contract records the
rank defect or excess and the molecular state must be reconsidered.

The novelty of \SONIC\ lies in the constraints imposed on this basis selection.
The accepted set \(\mathcal{A}\) is not just any basis of
\(\mathcal{R}(\mathcal{P})\).  It is a chemically ordered basis selected under a
source-family preorder: special fragment, center and ring coordinates are
tested before ordinary rank-completion candidates, and typed chemical families
are reduced before unrelated families are allowed to compete.  This
replaces the search for an arbitrary non-redundant basis by the construction of
a basis with explicit semantic priorities.

\section{Construction Principles}

The construction is easiest to state as a sequence of constraints on the
candidate space.

\begin{enumerate}
\item All covalent bonds in the validated topology have stretch rows.
\SMITH\ generates them in minimal mode or validates and classifies supplied candidates.
Stretches establish the bonded frame and are not allowed to be displaced by
later valence or torsional candidates.
\item Ordinary local valence-angle and linear-bend rows are generated or
supplied at the same boundary.  \SMITH\ groups them around each bonded center.
When a center has equivalent ligands or high coordination, the candidate angles
are grouped by local equivalence or by recognized idealized coordination
templates before block-local reduction.  Named templates are optional: every
coordination number above four has a template-independent equivalence fallback.
In the optional local-pseudosymmetry
path, equivalent stretch and angle orbits are replaced by a canonical
orthonormal SALC block with the local \(A_1\) row first.
\item Exocyclic torsions are generated only for non-ring central bonds with
non-terminal substituents.  The default intramolecular path keeps one
representative torsional coordinate per exocyclic bond, selected by a CIP-like
priority based on the continuous effective atomic numbers and local
connectivity.  Symmetry-equivalent alternatives are averaged when they belong
to the same local orbit.
\item Endocyclic valence-angle and curvilinear out-of-plane source spaces are generated ring
by ring.  Their active combinations are not arbitrary lists of primitive
angles or dihedrals; they are block-local deformation and puckering components.
For an \(n\)-membered ring, the default active puckering source dimension is
\(n-3\), with Fourier-style component vectors and optional
bond-order-based flexibility factors.  Ring stretches, valence angles and
out-of-plane puckering sources remain separate local source blocks, and the cyclic operations used
for local ordering must preserve the atom/synthon coloring of the ring.
\item Fused and condensed rings introduce additional torsional sources,
including butterfly coordinates around bonds shared by two rings.  These rows
remain in ring/torsional source blocks during reduction and symmetry
adaptation; they are not relabelled as generic torsions.
\item Special coordinates--fragment-center distances, center-atom distances,
fragment translations, fragment orientations and interaction-center distances--
are generated before ordinary candidates fill the remaining rank.  Their rows
receive priority because they describe objects that ordinary valence coordinates
can span algebraically but not semantically.
\item The final non-redundant basis is selected by analytic \(\Bmat\)-row rank,
using a fixed family order and a priority-first policy.  Symmetry is applied
after a valid non-redundant basis exists, and only inside typed source
blocks.
\end{enumerate}

This ordering differs from a global diagonalization of the full primitive
metric, whether expressed as \( \Bmat \Bmat^T \), as a row-space projector such
as \( \Bmat \Bmat^+ \), or as a single SVD of every redundant primitive at
once.  Such decompositions are mathematically useful, but their eigenvectors
depend on how the primitive rows have been scaled.  A stretch measured in
angstrom, a bend measured in radians, a torsion, a fragment translation and a
center distance have different natural units and different chemical roles.
Changing their weights can change which delocalized coordinates are retained.
\SONIC\ instead uses rank-revealing linear algebra inside chemically typed
blocks and reserves the final cross-block competition for a fixed
analytic \(\Bmat\)-rank test.  This does not make the construction unique in an
abstract mathematical sense.  It replaces hidden numerical weighting choices by
an explicit chemical priority order whose provenance is stored in the frozen
definition.

\section{Primitive Source and SMITH Coordinate Families}

The primitive source layer is intentionally redundant.  Redundancy is useful at this
stage because it allows chemical rules to create all plausible local
deformations before the rank selector removes dependent rows.  The important
restriction is that source primitives receive SMITH family labels that survive
reduction, symmetry adaptation and output.

\subsection{Bonded Coordinates}

Every covalent edge supplies one stretch. Equivalent stretches can be converted
to local sums and differences using a block-local SVD of their analytic
\(\Bmat\) rows, but the source block remains a stretch block.  Ordinary bends
are generated from pairs of neighbors around each center.  Nearly linear
centers produce two linear-bend components, because one ordinary angular
derivative becomes singular at linearity.

For three- and four-coordinate centers, \SONIC\ uses local equivalence classes to
construct chemically recognizable bend combinations: symmetric deformations,
rocking coordinates and tetrahedral or \(WXY_3/W_2XY_2\)-like patterns.  For
coordination numbers between five and nine, the construction first attempts to
match an idealized local template.  For every coordination number greater than
four, including those outside the named-template range, ligand-equivalence
classes provide a general fallback.  This is the practical
realization of the ``\(XY_n\) local symmetry'' idea: the algebra is local, the
grouping is chemically typed, and the result does not mix angle rows with
stretches or torsions.

\subsection{Local Pseudosymmetry and Canonical SALCs}

The molecular point group can be \(C_1\) even when a methyl group, a local
coordination sphere or a ring retains a useful approximate symmetry.  \SMITH\
therefore provides an optional local-pseudosymmetry construction before the
exact molecular projector.  For a bonded center, non-ring ligands are first
partitioned by the continuous effective atomic number or synthon descriptor
and by radial shell.  The current validated defaults are
\(5\times10^{-4}\) for the effective-atomic-number descriptor and
\(10^{-3}\)~\AA\ for the center--ligand distance.  These are construction
tolerances, not physical constants; they can be selected in the CLI or GUI and
are written to the diagnostic record.
The multiplicity pattern and ligand geometry then identify the local template,
for example \(C_{3v}\) for a \(WXY_3\) center or \(T_d\) for \(XY_4\).
Coordination numbers 5--9 are treated by the same path.  The current template
pairs are trigonal-bipyramidal/square-pyramidal (CN 5),
octahedral/trigonal-prismatic (CN 6), pentagonal-bipyramidal/capped-octahedral
(CN 7), square-antiprismatic/dodecahedral-like (CN 8), and
tricapped-trigonal-prismatic/capped-square-antiprismatic (CN 9).  If no
template is available, the best score is outside threshold, or two scores are
too close, the construction falls back to ligand-equivalence classes without
changing the final rank criterion.  If the sorted vectors of the
\(M=n(n-1)/2\) ligand-pair cosines are \(\mathbf c\) and
\(\mathbf t_T\), the template score is
\begin{equation}
 s_T=\left[\frac{1}{M}\sum_{k=1}^{M}(c_k-t_{T,k})^2\right]^{1/2}.
 \label{eq:local-template-score}
\end{equation}
Sorting makes this score invariant to Cartesian orientation and ligand order.
The best template is accepted only if \(s_1\le s_{\max}\) and
\(s_2-s_1\ge\Delta_s\); the defaults are \(s_{\max}=0.12\) and
\(\Delta_s=0.02\).  The selected template, best score, margin, decision status
and thresholds are frozen in the \SONIC\ contract.  The same coordinate
identity is therefore retained in every use of the contract rather than
reclassifying the center at each evaluation.
The diagnostic additionally stores the signed headroom from both acceptance
boundaries.  Assignments within ten percent of a selected threshold are marked
as near the RMS boundary, near the margin boundary, or near both.  This warning
does not rebuild the basis; it makes a potentially sensitive local decision
visible before a geometry trajectory is started.
Angles are partitioned separately according to the pair of ligand classes and,
for an accepted high-coordination template, the ideal pair-cosine class, so
that the SALC construction never mixes a stretch orbit with an angle orbit or
two chemically different angular blocks.

For an ordered orbit of \(m\) equivalent primitive rows
\(p_1,\ldots,p_m\), the first stored combination is
\begin{equation}
  q_{A_1}=\frac{1}{\sqrt{m}}\sum_{i=1}^{m}p_i .
\end{equation}
The remaining deterministic Helmert rows are
\begin{equation}
  q_{k+1}=\frac{\sum_{i=1}^{k}p_i-kp_{k+1}}{\sqrt{k(k+1)}},
  \qquad k=1,\ldots,m-1.
  \label{eq:local-helmert-salc}
\end{equation}
They form an orthonormal complement to \(q_{A_1}\), have zero coefficient sum
and fix a reproducible printed gauge.  Unless a full local character table has
also been resolved, the implementation deliberately labels these rows
\texttt{NON\_A1\_1}, \texttt{NON\_A1\_2}, \ldots rather than assigning an
unsupported local irrep.  Thus ``local \(A_1\)'' is an ordering and partitioning
property; the molecular irrep remains the one assigned later by the exact
point-group projector.

For a ring, the candidate local operations are the rotations and reversals of
the cyclic atom order.  Only operations preserving the effective-atomic-number/
synthon coloring and the outgoing-bond signature are retained.  The resulting
cyclic subgroup is recorded with its operation count and confidence, while
ring stretches, endocyclic valence angles and ring out-of-plane sources are transformed in
separate blocks.  In a fused system a shared edge is owned by the first ring
stretch domain and reported as shared in the subsequent domain, avoiding a
duplicate stretch source.  For an exocyclic central bond, equivalent primitive
dihedrals are combined as one local-rotor \(A_1\) orbit only when the signed
analytic Wilson row has nonzero rank, adequate norm stability and a continuous
local reference branch.  Otherwise the existing one-dihedral rule is used and
the failed rank/stability test is recorded.  Every generated row still passes the final analytic
\(\Bmat\)-rank test.

\subsection{Exocyclic Torsions}

For an acyclic bond \(j-k\), all substituent pairs \(i-j-k-l\) are possible
primitive torsions.  Keeping every such torsion is usually unnecessary and can
create symmetry noise.  The default intramolecular path therefore uses the
one-dihedral rule: among non-linear substituent pairs, the selected pair is the
highest-priority pair under effective-atomic-number and local-connectivity
ordering.  If several pairs belong to the same local equivalence orbit, the
coordinate is the normalized average of the corresponding primitive dihedrals.
This gives one chemically stable torsional coordinate for each eligible
exocyclic bond while still respecting local degeneracy.
Here ``average'' means a stored linear combination of primitive torsional
Wilson rows, together with a scalar value on the branch fixed by the reference
geometry; it is not an arithmetic mean of periodic angles across a
\(360^\circ\) wrap.  When a finite torsional value is needed for reporting or
for a nonlinear model coordinate, \SONIC\ uses the local reference branch, or an
explicit circular representation through \(\sin\phi\) and \(\cos\phi\).  Thus
symmetry averaging never identifies \(179^\circ\) and \(-179^\circ\) with an
unphysical zero-degree torsion.

\subsection{Rings, Puckering and Butterfly Coordinates}

Ring coordinates are the main place where primitive internal coordinates must
be replaced by natural combinations.  An \(n\)-membered ring has redundant
endocyclic angle and out-of-plane source lists.  \SONIC\ constructs ring deformation
coordinates and \(RPck\) puckering components as linear combinations of those
source rows, keeping \(n-3\) independent puckering components for rings larger
than three atoms.  The endocyclic valence-angle combinations are the CyGNA
source family and are stored as \(RDef\) rows; the corresponding endocyclic
out-of-plane combinations are stored as \(RPck\) rows.  For each ordered ring
stencil \(D(a,b,c,d)\), the native source is the curvilinear coordinate
\(U(b,a,d,c)\); the dihedral only specifies the cyclic atom ordering.  These
signed U sources are the closest local curvilinear counterpart of the
Cremer--Pople puckering displacements, while retaining analytic Wilson rows
and a direct local definition for planar and nonplanar rings.  This is a
geometric analogy, not an identification: Cremer--Pople coordinates describe
global ring-shape amplitudes and phases, whereas the (U) rows are local,
signed curvilinear sources from which the active (RPck) combinations are
constructed.  The polar
Cremer--Pople-like descriptors \(Q\) and \(\phi\) are useful for
interpretation, but the active coordinate basis stores linear \(RPck\) rows
because their \(\Bmat\) rows are analytic linear combinations of out-of-plane
derivatives.\cite{CremerPople1975}
For every selected cosine--sine puckering pair, \SMITH\ can also emit the
derived observables
\begin{equation}
\label{eq:puckering-polar-observables}
  QPck_m=\sqrt{(RPck_m^c)^2+(RPck_m^s)^2},\qquad
  \Phi P_m=\operatorname{atan2}(RPck_m^s,RPck_m^c).
\end{equation}
These \(QPck/\Phi P\) variables are functions of the stored linear rows rather
than additional rank-carrying coordinates.

The ring rows are a constructive extension of the natural-coordinate principle
to cyclic source lists rather than a direct quotation of a published
molecule-specific formula.  If \(s_j\) denotes such a list, for example the
endocyclic valence angles or a curvilinear out-of-plane puckering list, the stored rows are
real Fourier-like combinations
\begin{equation}
\label{eq:ring-fourier-combinations}
  Q_m^c = N_m\sum_{j=1}^{n} s_j\cos\frac{2\pi m(j-1)}{n},\qquad
  Q_m^s = N_m\sum_{j=1}^{n} s_j\sin\frac{2\pi m(j-1)}{n},
\end{equation}
with the alternating row \(Q_{n/2}=n^{-1/2}\sum_j(-1)^{j-1}s_j\) for even
\(n\).  The normalization \(N_m\) is chosen so that coefficient vectors are
unit length before the symmetry projection.  Thus the same cyclic algebra is
used for the CyGNA valence-angle family and for \(RPck\) out-of-plane puckering
rows; only the source primitive list changes.

For a six-membered ring segment \(1\cdots6\), a typical generated angle row has
the explicit form
\begin{equation}
\label{eq:explicit-ring-rdef}
\begin{aligned}
RDef &=
 0.57735\,A(6,1,2)-0.28868\,A(1,2,3)-0.28868\,A(2,3,4) \\
&\quad +0.57735\,A(3,4,5)-0.28868\,A(4,5,6)-0.28868\,A(5,6,1),
\end{aligned}
\end{equation}
and the analogous puckering row is built from the corresponding cyclic
\(U\) coordinates with the same normalized coefficient pattern.  In the frozen
contract these rows carry family labels, source primitive identifiers and
symmetry labels; \(Q\) and \(\phi\) are then functions of the \(RPck\) rows, not
replacement active coordinates.
Table~\ref{tab:ring-coordinate-examples} gives representative generated ring
rows in a compact cyclic notation.  Let
\(A_i=A(i-1,i,i+1)\) and
\(u_i=U(i+1,i,i+3,i+2)\), with atom indices taken
modulo six.  The printed signs are one valid gauge; for degenerate irreps a
different orthonormal pair spans the same symmetry subspace.

\begin{table}[htbp]
\centering
\small
\caption{Representative six-membered-ring rows generated by \SONIC.  The
examples show the kind of coordinate stored in the contract rather than only
the family count.  A symmetry-preserving optimization or semiexperimental fit
may request only the totally symmetric subset, e.g. the \(A_{1g}\) rows for the
\(D_{3d}\) example, while GF/PED, DVR and full-rank checks can retain the full
symmetry-labelled basis.}
\label{tab:ring-coordinate-examples}
\begin{tabular}{@{}p{0.20\linewidth}p{0.14\linewidth}p{0.58\linewidth}@{}}
\toprule
Stored row & Symmetry & Linear form \\
\midrule
\(A_{1g}RDef001\) & \(A_{1g}\) &
\(\left(2A_1-A_2-A_3+2A_4-A_5-A_6\right)/\sqrt{12}\) \\
\midrule
\(A_{1g}RPck001\) & \(A_{1g}\) &
\(\left(2u_1-u_2-u_3+2u_4-u_5-u_6\right)/\sqrt{12}\) \\
\midrule
\(E_uRPck001\) & \(E_u\) &
\(\left(u_1-u_2-u_4+u_5\right)/2\) \\
\midrule
\(E_uRPck002\) & \(E_u\) &
\(\left(u_1+u_2-2u_3-u_4-u_5+2u_6\right)/\sqrt{12}\) \\
\bottomrule
\end{tabular}
\end{table}

Unsaturated rings and locally rigid ring fragments require more than a generic
\(n-3\) puckering count.  The reconstruction method of Lema-Saavedra and
Fern{\'a}ndez-Ramos explicitly reduces the puckering dimensionality in the
presence of rigid or multiple bonds and then recovers the full ring geometry by
distributing the remaining distortions under constraints.\cite{LemaSaavedraFernandezRamos2026RingCP}
The present \SONIC\ construction already introduces bond-order-dependent
flexibility factors in \(RPck\) source vectors, so out-of-plane sources
associated with more rigid ring bonds contribute less to the puckering row.
The rigid-bond treatment of
Ref.~\citenum{LemaSaavedraFernandezRamos2026RingCP} is therefore most naturally
viewed in \SONIC\ as a stronger, optional source-space contraction rather than as
a rewrite of the covalent graph.  A rigid bond \(i-j\) can be replaced, inside
that ring source polygon only, by the virtual source point
\(\mathbf{c}_{ij}=(\mathbf{r}_i+\mathbf{r}_j)/2\); the effective polygon then
has one fewer source point and one fewer puckering degree of freedom.  The
analytic \(\Bmat\) row is expanded back to the real atoms by the chain rule,
with the bond-center derivative distributed over \(i\) and \(j\).  This
contraction is attractive for genuine double bonds, conjugated fragments and
locally stiff fused-ring units, but it must be applied to complete symmetry
orbits and only when the molecular-state continuous bond-order and synthon
descriptors mark a hard local constraint.  The present bond-order flexibility factors are
therefore the conservative default; rigid-source contraction is the planned
extension for cases where the chemically intended model removes, rather than
merely down-weights, a non-puckering motion from the active source space.  In
both cases, the active GIC basis remains linear while Cremer--Pople amplitudes
and phases are exposed as derived functions for sampling, PES fitting and
conformer classification.

Fused, spiro and bridged systems are handled as connected ring-source spaces
rather than as ordinary acyclic torsions.  When two rings share a bond, \SONIC\
adds butterfly coordinates around that shared bond.  Condensed-ring torsions
and ring puckering components remain separate homogeneous blocks in the
symmetry projector.  This is essential for high-symmetry systems such as cubane
and for fused aromatic systems where a global torsion list would obscure the
local ring role.

\subsection{Out-of-Plane and Improper Coordinates}

Three-coordinate centers generate native analytic out-of-plane coordinates
unless all four atoms involved are cyclic.  This \(U\) definition is used both
for the primitive source and for the final SONIC coordinate; an improper
dihedral is never selected during primitive generation or SMITH reduction.  Only
the Gaussian~16 \texttt{.gjf} writer translates
\(U(I,J,K,L)\mapsto D(J,I,L,K)\).  Thus Gaussian compatibility is a terminal
serialization step and cannot change the native coordinate model.

\section{Rank Reduction}

The rank-reduction step takes the ordered primitive set
\(\mathcal{P}=(p_1,\ldots,p_m)\), the analytic rows
\(\bmat(p_i)\) and a source-family map \(F:\mathcal{P}\to\mathfrak{F}\).
The family set \(\mathfrak{F}\) carries a priority preorder
\(\preceq_{\mathfrak{F}}\).  Priority families--fragment coordinates,
interaction-center distances, ring-puckering sources and related special
objects--are minimal elements in this preorder; ordinary stretches, bends and
torsions fill the rank only after those rows have been tested.  Ties inside
a family are resolved by topology, continuous-perception equivalence and
CIP-like priority keys stored in \(\mathcal{M}\).  Thus the construction works on a
fixed total order that extends a chemically motivated partial order.

The following design invariants are consequences of that ordered protocol.
They are stated explicitly to make the construction verifiable, not to claim
uniqueness over all possible internal-coordinate bases.

\paragraph{Relabelling covariance.}
\label{thm:relabel}
Let two frozen molecular states be related by an atom relabelling \(\pi\) that
preserves element identities, geometry, topology, perception descriptors, fragments,
interaction centers, point-group operations, priority keys and tolerances.  Let
\(Q_\pi\) be the induced orthogonal permutation of Cartesian displacement
coordinates and \(U_\pi\) the induced permutation/sign map on primitive
coordinate rows.  If \SONIC\ returns a contract \(\mathcal{C}\) for the first
state, then it returns an isomorphic contract \(\pi_*\mathcal{C}\) for the
second state.  In particular,
\[
  \Bmat(\pi_*\mathcal{P})
  =
  U_\pi \Bmat(\mathcal{P}) Q_\pi^T,
\]
and the accepted coordinate spaces, family labels and symmetry-adapted blocks
are carried into each other by the induced row and coefficient-space
isomorphisms.
This follows because every primitive generator, row norm, rank residual and
homogeneous-block projector is evaluated from data preserved by \(\pi\);
orthogonal permutation/sign transformations do not change the Boolean
acceptance tests.

Equivalently, \SONIC\ is natural with respect to symmetry-preserving atom
relabellings: relabelling the molecular state and then constructing the
contract gives the same object, up to the canonical contract isomorphism, as
constructing the contract first and then transporting it by the relabelling.

\paragraph{Reproducible contract.}
\label{prop:determinism}
For a fixed molecular state--geometry, topology, perception descriptors, point-group
operations and fragment data--and for fixed primitive generators, family
priority order, tie-breaking keys, row-normalization rule and numerical
tolerances, \SONIC\ returns a unique contract, or a unique
incompatibility certificate.  In particular, no stochastic search, optimizer
history or manual choice is part of the contract definition.
The verification is direct: each stage is a fixed map of the frozen
input data, and the fixed order of candidate rows converts every rank and
symmetry decision into a stored threshold test.

\paragraph{Semantic preservation.}
\label{prop:protected}
Let \(\mathcal{P}_{\mathrm{prot}}\) be the priority prefix of the ordered
candidate set.  Every candidate in this prefix whose Wilson row is nonsingular
and independent, within the stored tolerance, of the previously accepted prefix
rows is retained in any returned final object.  No ordinary stretch, bend or
torsion can displace such an accepted coordinate.  If the independent prefix
already exceeds the target vibrational rank, \SONIC\ reports an invalid model
rather than removing a semantic coordinate by exchange.  This follows directly
from the ordered-prefix construction.

\paragraph{Rank-preserving selection.}
\label{prop:rank-selection}
Assume exact arithmetic and define
\[
  \mathcal{R}_k=\operatorname{span}\{\bmat(p_1),\ldots,\bmat(p_k)\}.
\]
Incremental Gram--Schmidt selection accepts \(p_k\) exactly when
\(\bmat(p_k)\notin\mathcal{R}_{k-1}\).  After all candidates have been tested,
the accepted subset \(\mathcal{A}\) is an ordered basis of
\(\mathcal{R}(\mathcal{P})\).  If
\(\operatorname{rank}\Bmat(\mathcal{P})=r\), then
\(|\mathcal{A}|=r\) and
\(\operatorname{span}\Bmat(\mathcal{A})=\operatorname{span}\Bmat(\mathcal{P})\).
Indeed, at step \(k\) Gram--Schmidt decomposes \(\bmat(p_k)\) into its
projection on the current accepted span plus an orthogonal residual; a zero
residual leaves the span unchanged, and a nonzero residual adds exactly one
new independent direction.

The numerical construction is the finite-precision version of this statement.  \SONIC\
selects rows by incremental modified Gram--Schmidt rank testing with a fixed
numerical tolerance, but the comparison is performed on normalized Wilson rows,
\begin{equation}
  \hat{\bmat}(p) = \frac{\bmat(p)}{\|\bmat(p)\|_2}.
\end{equation}
The residual tested for candidate \(p\) is
\begin{equation}
  \rho(p) =
  \left\|
    \hat{\bmat}(p) -
    \sum_{i \in \mathcal{A}}
    \left[\hat{\bmat}(p)\cdot \hat{\bmat}_i\right]\hat{\bmat}_i
  \right\|_2 ,
\end{equation}
where \(\mathcal{A}\) is the current accepted orthonormal row basis.  The
absolute row norm is used only to reject singular coordinates; independence is
decided by the direction of the analytic Wilson row.  This normalization avoids
a direct comparison of raw angstrom and radian scales in the final rank test.
It does not remove all modelling choices: the ordered candidate list remains a
chemical convention, not a uniqueness guarantee.  Singular rows and rows
dependent on previously accepted rows are recorded in the frozen diagnostics.

The decisive policy is therefore not only the linear algebra routine; it is the
priority constraint under which the basis is built.  Special coordinates are
tested before ordinary primitives.  Ordinary stretches, bends and torsions then
complete the rank.  A priority coordinate can be skipped only if it is singular
or dependent on previously accepted priority rows.  It cannot be displaced
merely because an ordinary valence coordinate spans a similar first-order
motion.  If these rows alone exceed the vibrational rank, the construction is
rejected rather than silently discarding a special coordinate.

\paragraph{Block-local scaling invariance.}
\label{thm:block-scaling}
Let \(H\subset\mathcal{P}\) be a homogeneous coordinate block.  Uniformly
rescaling all Wilson rows in that block by a nonzero scalar does not change the
normalized-row rank decisions inside \(H\), nor the singular-vector or projector
subspaces obtained from \(\Bmat_H\Bmat_H^T\).  By contrast, a global
decomposition of several heterogeneous blocks is generally not invariant to
independent rescalings of those blocks.
The reason is that row normalization removes a uniform factor inside \(H\), and
\((\lambda\Bmat_H)(\lambda\Bmat_H)^T=\lambda^2\Bmat_H\Bmat_H^T\) rescales
eigenvalues but not eigenspaces; concatenating heterogeneous blocks before
diagonalization changes their relative weights.

\paragraph{Local stability.}
\label{thm:local-stability}
Let \(S_0\) be a nonsingular frozen molecular state.  Assume that, in a
neighbourhood of \(S_0\), topology, fragment decomposition, perception equivalence
classes, point-group representation, priority prefix, typed block
membership and priority order remain fixed in a ball of radius
\(\delta_{\mathrm{str}}\) around \(S_0\).  Let
\(\{f_a(\x)\}_{a=1}^{n_t}\) be the finite set of scalar decision functions used
by the construction in this stratum: row norms, rank residuals, block singular
or projected-vector norms and any sign-gauge pivot quantities used to store
SALC coefficients.  Let \(\tau_a\) be the corresponding acceptance threshold and
define the margin at \(S_0\)
\[
  \mu_a = |f_a(\x_0)-\tau_a| > 0 .
\]
Assume that each \(f_a\) is Lipschitz on the whole stratum ball with a uniform
constant \(L_a\), i.e.,
\[
  |f_a(\x)-f_a(\y)| \le L_a\|\x-\y\|
\]
for all \(\x,\y\) in that ball.  Then every
\[
  0 < \delta <
  \min\left[
    \delta_{\mathrm{str}},
    \min_{a:L_a>0}\frac{\mu_a}{2L_a}
  \right]
\]
preserves the discrete contract, with the usual convention that the
inner minimum is \(+\infty\) if all \(L_a=0\).  Thus every state \(S\) in the
same discrete stratum with \(\|\x(S)-\x(S_0)\|<\delta\) generates an isomorphic
contract: the primitive identities, accepted/rejected rank decisions,
priority coordinates, family labels and symmetry labels are unchanged, while
\(\Bmat\) rows and coefficient matrices vary continuously with the geometry
under the same normalization convention.
This is the usual finite-decision margin argument: each decision function stays
on the same side of its threshold for perturbations smaller than
\(\mu_a/(2L_a)\), and the minimum over the finite decision set preserves the
entire ordered construction.

This is the technical reason for block-local reduction.  \SONIC\ does not claim
that weights disappear from all possible coordinate constructions.  It avoids
using arbitrary cross-family weights to define chemically labelled variables.
Only after typed blocks have been reduced and special coordinates have
been accepted does the final normalized rank audit compare directions across
families.

The local-stability result is deliberately stratum-local rather than global.  A
change in topology, symmetry, local equivalence or a near-threshold residual is
a genuine model change; it is therefore reported as contract provenance rather
than hidden behind a re-diagonalized global basis.

This policy is what makes the coordinate set suitable for refinement.  A
fragment-center distance or ring-center to atom distance might be expressible as
a linear combination of many ordinary internal derivatives near a reference
geometry.  That algebraic fact does not make the ordinary combination a useful
fit variable.  The priority-first rule preserves the variables that define the
model, then uses ordinary coordinates for completeness.

\section{Symmetry Adaptation}

Two symmetry levels are kept distinct.  The local construction described above
uses center, ring or bond domains to obtain chemically ordered source blocks
even for a globally asymmetric molecule.  Exact molecular symmetry is applied
only after a non-redundant coordinate basis has been selected and remains the
final authority for molecular irrep labels and totally symmetric export.  Let
\(G_{\mathrm{pt}}\) be the molecular point group stored in the
source state.  The point group is not guessed from idealized element labels
alone: the molecular-state builder first clusters atoms by continuous element,
coordination, bond-order and synthon descriptors, then tests candidate
rotations, reflections and inversion operations against the Cartesian geometry
within the stored tolerances.  An operation is accepted only when it maps each
atom onto an atom of the same continuous equivalence class and preserves the
stored topology and local descriptor signatures.  The accepted operations, atom
permutation table, tolerances and perception version are frozen before
\SONIC\ constructs coordinates.  For a path whose symmetry changes, the user
may request a reduced working subgroup, for example \(C_1\), \(C_s\), \(C_i\)
or \(C_2\).  \SONIC\ then filters the stored operations to that subgroup and
uses the reduced group consistently for labels, projectors and totally
symmetric coordinate export.  Each operation \(h\in G_{\mathrm{pt}}\) is represented by a
Cartesian rotation matrix, an atom permutation and the signs required for
oriented coordinates such as torsions, out-of-plane bends and axial fragment
rotations.  For a homogeneous accepted coordinate block \(H\), with coefficient
space \(V_H\), the stored operations map the block into itself.  Since each
homogeneous block admitted to the projector is invariant under this permutation
and sign representation, its coefficient space carries a finite-dimensional
representation of the point group,
\[
  \Gamma_H:G_{\mathrm{pt}}\rightarrow GL(V_H).
\]
In the simplest case \(\Gamma_H(h)\) permutes equivalent coordinate basis
vectors and multiplies them by the appropriate parity sign.

If \(H\) is closed under the stored operations, the symmetry-adapted subspace of
irreducible representation \(\alpha\) is obtained by the standard character
projector
\begin{equation}
  \Pi_{\alpha}^{(H)}
  =
  \frac{\ell_\alpha}{|G_{\mathrm{pt}}|}
  \sum_{h\in G_{\mathrm{pt}}}
  \chi_\alpha(h)^*\,\Gamma_H(h),
\end{equation}
where \(\ell_\alpha\) is the dimension of \(\alpha\) and
\(\chi_\alpha\) its character.  \SONIC\ applies these projectors to the coefficient
space of the block, orthonormalizes the nonzero projected vectors and stores the
resulting SALC matrix \(C_{\alpha H}\).  The symmetry-adapted Wilson rows are
then
\[
  \Bmat_{\alpha H}=C_{\alpha H}\Bmat_H .
\]
For tasks that must preserve the stored point-group symmetry, \SONIC\ can export
only the totally symmetric rows, such as \(A_1\), \(A'\) or \(A_{1g}\)
depending on the group.  This is the natural subset for symmetry-preserving
geometry optimizations and semiexperimental structure fits.  The remaining
irreps are still retained in the full contract for rank closure, GF/PED block
analysis, DVR coordinates and diagnostics.
If an irreducible representation occurs with multiplicity greater than one in
the same homogeneous block, the projector defines a subspace, not a unique
printed basis.  \SONIC\ fixes the stored basis by applying modified
Gram--Schmidt to projected source vectors in the canonical block order, by
discarding vectors below the saved norm tolerance, and by choosing the sign from
the largest-magnitude coefficient pivot.  The subspace is the representation
theoretic object; the ordered, sign-gauged coefficient matrix is the
deterministic serialization stored in the contract.
Along scans or optimizations, a change in the identity of the largest pivot can
flip the printed gauge even when the invariant subspace, rank and symmetry
labels are unchanged.  \SONIC\ therefore treats pivot changes as executable path
diagnostics: consecutive SALC coefficient matrices are compared by subspace
overlap, pivot changes are flagged as gauge events, and the same diagnostic
module provides a Procrustes alignment for downstream GF/PED or DVR layers
when a path-continuous row basis is required.  None of the reported
chair-to-boat, ferrocene or GF/PED probes showed a pivot switch in the active
blocks used for the tabulated quantities.

The block-local rule is the key safeguard.  A stretch and a torsion may
belong to the same irreducible representation, but they are not the same kind
of coordinate.  Mixing them during symmetry adaptation would produce a valid
vector in an abstract representation and an opaque variable in a chemical
model.  \SONIC\ therefore projects only within coordinate families and source
blocks.  If a block is not closed under the stored operations, the construction uses
an explicit local SALC fallback only where the contract allows it, and records
that fallback in the symmetry diagnostics.

Fragment translations transform as polar vectors; fragment orientations
transform as axial vectors.  Torsions and out-of-plane coordinates carry the
appropriate sign changes under atom permutation and inversion.  Ring puckering
coordinates are projected through the underlying dihedral source vectors so
that the symmetry labels refer to the same analytic \(\Bmat\) rows used later
by GF, fitting and optimization.

\section{Intermolecular and Special Coordinates}

Weak complexes are a stress test for any internal-coordinate generator.  A
covalent graph alone is disconnected, but adding arbitrary weak contacts as
ordinary bonds can create artificial rings and mix covalent and non-covalent
motions.  Within its current validated scope, \SONIC\ therefore keeps two
explicit intermolecular constructions: a fragment-coordinate model
and a pseudo-bond/pseudo-cycle model for selected hydrogen-bonded or
contact-closed graphs.

In the release-grade \texttt{SPECIAL\_COORDINATES} mode, each fragment keeps its
intramolecular coordinate basis.  Interfragment motion is represented by
special coordinates: fragment-center distances, fragment-center to
atom distances, Cartesian components of relative fragment translation and
relative fragment orientations expressed as exponential-map rotation
coordinates.  This construction explicitly adopts the fragment-rotation
parametrization introduced by Wang and Song in TRIC/geomeTRIC: collective
translations are described through fragment centroids, while orientations are
represented by an exponential map of quaternions.\cite{WangSong2016GeomeTRIC}
The use of this exponential mapping in \SONIC\ should therefore be credited to
the geomeTRIC work.

This shared parametrization does not make the two coordinate constructions
identical.  In TRIC/geomeTRIC, three translations and three rotations are added
for each fragment and are subsequently combined with the intrafragment
primitives in a delocalized coordinate space.\cite{WangSong2016GeomeTRIC}
Within the \SONIC\ contract, one fragment is instead selected deterministically
as the reference; three relative translations and, for non-linear partners,
three relative exponential-map orientations are generated for every other
fragment, so that the overall rigid translation and rotation of the complete
complex do not enter the molecular vibrational contract.  These rows remain a
protected semantic block during rank reduction rather than being delocalized
with stretches, bends and torsions.  They transform explicitly as polar and
axial vectors under the stored point-group operations and are serialized
together with their centers, body frames, gauge choices, symmetry labels and
analytic \(\Bmat\) rows.  Thus \SONIC\ adopts and credits the TRIC exponential
map while embedding it in a distinct, frozen coordinate construction.

Fragment coordinates are not the only intermolecular representation available
in \SONIC.  For a weak complex, the coordinate contract can select either the
\texttt{SPECIAL\_COORDINATES} fragment construction just described or the
\texttt{PSEUDO\_BONDS} construction.  The former preserves the disconnected
monomers and represents their relative rigid-body motion explicitly; the latter
inserts selected weak contacts as typed pseudo-edges and generates the
associated pseudo-cycle coordinates without reclassifying those contacts as
covalent bonds.  The selected construction is recorded in the contract, so the
two descriptions remain explicit alternatives rather than indistinguishable
members of one primitive pool.

Fragment centers, body frames and orientation-chart references are serialized
with the coordinate definition.  Relative orientations use the exponential-map
chart on \(\mathrm{SO}(3)\), with the same finite-rotation parametrization as
TRIC/geomeTRIC, but SMITH does not prescribe how a requested finite coordinate
change is converted back to Cartesian geometry.  Choice of right inverse,
damping, finite rotations, continuation, chart rebasing, ``soft'' and ``hard''
partitions, symmetry maintenance and backend-frame alignment all belong to the
application using the contract.  The SMITH motion files discussed below are
explicitly first-order visualizations at fixed \(\Bmat\), not general
internal-to-Cartesian solvers.

Interaction centers extend the same idea to non-atomic reference objects.  A
center may be a ring centroid, a bond midpoint or a contact center.  Distances
from atoms or fragments to such centers are special coordinates and
are evaluated analytically by distributing the center derivative over the atoms
that define it.  This gives a practical route to coordination complexes,
metallocenes and eta-like interactions without pretending that a metal-ring
centroid is a covalent atom in the molecular graph.

Ferrocene provides a compact example.  If the two cyclopentadienyl ring
centers are
\[
 C_{031}=\frac{1}{5}\sum_{i\in R_1}\mathbf{x}_i,\qquad
 C_{032}=\frac{1}{5}\sum_{i\in R_2}\mathbf{x}_i,
\]
and the iron atom is atom 2, \SMITH\ first defines the two center--atom
distances
\[
 d_1=|C_{031}-\mathbf{x}_2|,\qquad d_2=|C_{032}-\mathbf{x}_2|.
\]
The protected symmetry-adapted rows retained in the contract are then
\[
 A_1'CnAt001=\frac{d_1+d_2}{\sqrt2},\qquad
 A_2''CnAt001=\frac{d_1-d_2}{\sqrt2}.
\]
The \(A_1'\) row is the total-symmetric metal--ring breathing coordinate; the
\(A_2''\) partner is kept in the full contract and marked frozen when only
total-symmetric coordinates are requested for an optimization or
semi-experimental fit.  In Gaussian syntax the centers are serialized as
inactive Cartesian functions such as
\[
 CxC031=(X_1+X_3+X_6+X_5+X_4)/5,
\]
and the active distance row is written as the corresponding square-root
function of the center coordinates and the iron Cartesian coordinates.  The
special coordinate is therefore generated from the molecular model, not
hand-entered as an extra atom.

The $\eta^3$ allyl--palladium standalone probe exercises the same interface with a
non-ring ligand center.  For allyl carbon atoms 3--5, the input state supplies
\[
 C_{001}=\frac{\mathbf{x}_3+\mathbf{x}_4+\mathbf{x}_5}{3}
\]
with center kind \texttt{ETA3\_CENTER}; \SMITH\ then protects
\[
 q_{\eta^3}=|C_{001}-\mathbf{x}_{\rm Pd}|.
\]
The idealized ten-atom fixture closes at rank \(24/24\), and the reference
value is \(q_{\eta^3}=1.8000\)~\AA.  Its Gaussian 16 export writes the three
center components as inactive Cartesian functions and the metal--center
distance as an active square-root expression.  This is deliberately an
interface and rank-closure test rather than an optimized electronic-structure
benchmark: definition of the $\eta^3$ center belongs to the input provider,
whereas its retention, first differentiation and serialization belong to \SMITH.

In the current validated \texttt{PSEUDO\_BONDS} baseline, selected
non-covalent contacts are inserted into the construction graph as typed
pseudo-edges, and the resulting pseudo-cycles generate angular and torsional
source components analogous to covalent ring coordinates.  These rows are still
labelled separately from covalent stretches, bends and torsions, so downstream
users can decide whether they are model coordinates, constraints or
diagnostics.  The present validation uses formic-acid--water, formamide--water
and the glycine I/II hydrogen-bond fixtures as regression probes: a single
intramolecular bridge and a bifurcated bridge both close the expected rank
without converting the underlying covalent topology into a different molecule.

A concrete non-covalent example is the formic-acid--water dimer used in the
regression suite.  In pseudo-bond mode, the validated weak contacts
\((3,8)\) and \((5,6)\) are inserted as typed \texttt{HBOND} pseudo-edges.
They are stored in the contract as
\[
  \texttt{PSEUDO\_BOND\_COUNT 2}, \qquad
  \texttt{1 3 8 KIND=HBOND}, \qquad
  \texttt{2 5 6 KIND=HBOND}.
\]
The augmented graph then produces ordinary stretch, bend and torsion candidates
plus protected pseudo-cycle bend and torsion families.  The final object
remains rank-complete at \(18/18\), and the pseudo-cycle rows remain labelled
separately from covalent ring coordinates so that downstream analyses can keep
or suppress them explicitly.

Additional pseudo-bond and intermolecular fixtures were added to keep this
baseline close to the weak-complex cases used by geomeTRIC-style
translation--rotation coordinates (Table~\ref{tab:pseudo-cycle-fixtures}).
The point of the table is deliberately modest: these systems verify that typed
contacts, fragment-center variables and pseudo-cycle rows can be constructed
and labelled without collapsing the fragment model into a covalent molecule.
They do not yet establish full generality for arbitrary non-covalent networks.

\begin{table}[htbp]
\centering
\scriptsize
\setlength{\tabcolsep}{3pt}
\renewcommand{\arraystretch}{0.95}
\caption{Additional weak-complex and pseudo-cycle fixtures used to define the
current validated baseline.  Nontrivial symmetry cases are included to test
that fragment and pseudo-bond rows can coexist with point-group perception.}
\label{tab:pseudo-cycle-fixtures}
\begin{tabular}{@{}p{0.20\linewidth}p{0.10\linewidth}p{0.24\linewidth}p{0.36\linewidth}@{}}
\toprule
System & PG & Coordinate mode & Diagnostic role \\
\midrule
Formic acid--water & \(C_s\) & typed H-bond pseudo-edges and fragment centers &
two weak contacts close a hydrogen-bonded pseudo-cycle while preserving the
monomer fragments. \\
\midrule
Water dimer & \(C_s\) & protected fragment translations, orientations and
center distances & geomeTRIC-like weak-complex rotor proxy; tests that
orientation rows have nonzero sparse derivatives in a non-covalent dimer. \\
\midrule
Li\(_2\)--Ar & linear/axial & protected interfragment axes plus polar and
elliptic functions & Gaussian-\GIC\ scan motif generated from fragment
translation axes rather than hand-written coordinate expressions. \\
\midrule
Borane--ammonia & \(C_{3v}\) & fragment-center distance and axial relative
orientation & nontrivial symmetric donor--acceptor complex; tests vector and
axial-vector transformation of fragment variables. \\
\midrule
Glycine I & \(C_1\) & intramolecular H-bond pseudo-edge & single internal
weak bridge; tests pseudo-cycle construction without intermolecular fragments. \\
\midrule
Glycine II & \(C_1\) & bifurcated intramolecular H-bond pseudo-edges &
multi-contact weak bridge; tests deterministic contact ordering and labelled
pseudo-cycle rows. \\
\bottomrule
\end{tabular}
\end{table}

\section{Validated Construction Scope}

Table~\ref{tab:status} records the status of the present \SONIC\
construction.  The purpose is not to document a program interface, but to state
which mathematical objects belong to the present contract and which
objects remain methodological extensions.  This separation is important because
downstream validation is meaningful only for coordinates that are part of the
frozen contract.

\begin{table}[p]
\centering
\scriptsize
\setlength{\tabcolsep}{2.6pt}
\renewcommand{\arraystretch}{0.88}
\caption{Validated scope of the main \SONIC\ coordinate families.}
\label{tab:status}
\begin{tabular}{@{}p{0.24\linewidth}p{0.17\linewidth}p{0.50\linewidth}@{}}
\toprule
Feature & Status & Notes \\
\midrule
Covalent stretches & Active & All topology bonds; local equivalence grouping available. \\
\midrule
Exocyclic bends & Active & Bonded-center bends; local symmetric combinations for common motifs. \\
\midrule
High coordination & Active & Two named templates per CN 5--9; generic ligand-equivalence fallback for every CN $>4$. Python is dynamically sized; the distributed Fortran backend is compiled through CN 32. \\
\midrule
One-dihedral exocyclic torsions & Active default & \texttt{ONEDIH} is the default intramolecular path; legacy alternatives are diagnostic. \\
\midrule
Ring deformation/puckering & Active & Endocyclic angle and torsion sources generate \(RDef\) and \(RPck\); \(Q/\phi\) are derived descriptors. \\
\midrule
Butterfly/condensed rings & Active & Fused-ring butterfly and condensed torsion families remain separate homogeneous source blocks. \\
\midrule
Out-of-plane rows & Active & Native \(U\) primitives and SONICs; improper dihedrals are generated only in a requested Gaussian 16 \texttt{.gjf} export. \\
\midrule
Special fragment coordinates & Active & Fragment-center distances, center-atom distances, translations and orientations are protected first. \\
\midrule
Interaction centers & Active baseline & Ring, bond and related virtual centers can enter center-atom coordinates. \\
\midrule
Semantic \texttt{PROTECT}/\texttt{OBSERVABLE} layer & Active baseline & Deduplication, row-equivalent substitution, rank-excess rejection and sparse observable propagation. \\
\midrule
Pseudo-bond/pseudo-cycle mode & Validated baseline & Typed weak contacts become pseudo-edges; pseudo-cycle bends/torsions stay labelled separately. \\
\midrule
Point-group projector & Active & Applied after non-redundant reduction inside homogeneous blocks; fallback is recorded. \\
\midrule
Local-pseudosymmetry SALCs & Active optional path & Center, ring and bond domains use effective-atomic-number/synthon and radial equivalence; the local \(A_1\) row is first and the exact molecular projector remains final. \\
\midrule
\SONIC\ symmetry-adapted Cartesian reference & Active & Cartesian active-space reference for validation; does not change \GIC\ construction. \\
\midrule
Sparse Wilson-\(\Bmat\) representation & Validated baseline & Sparse derivative rows are stored; dense matrices are only audit representations. \\
\midrule
First derivative of \(\Bmat\) & Outside the SMITH contract & \(\Bmat'\) is not required to construct SONIC and is neither built nor serialized by standalone SMITH.  A second-order consumer may evaluate it on demand from the frozen coordinate definitions. \\
\midrule
Primitive/function registry & Validated baseline & Canonical primitive and \(\y=f(\q)\) records prevent duplicate coordinate objects. \\
\midrule
Coordinate-function layer & Validated baseline & Inverse/exponential distances, trigonometric functions, polar radii and quadratic functions use sparse chain rules. \\
\bottomrule
\end{tabular}
\end{table}

The boundary at second order deserves an explicit statement.  The first
Wilson-matrix derivative is
\begin{equation}
 \mathcal D_{iab}=\frac{\partial B_{ia}}{\partial x_b}.
 \label{eq:sparse-b-first-derivative}
\end{equation}
This object is often denoted \(\Bmat'\), but it is not needed to define a
coordinate, to select the non-redundant SONIC basis or to evaluate its Wilson
matrix.  Accordingly, SMITH does not construct or store it.  A program that
must transform a Hessian at a nonstationary geometry is responsible for
evaluating \(\mathcal D\) on demand from the frozen primitive definitions and
SONIC coefficients.  Such an evaluator may assemble analytic primitive
Cartesian Hessians into sparse slices and use finite differences only as a
reported audit or fallback, but that is a second-order application service,
not part of the published SMITH interface.

The reason an application may need this service is the exact Hessian chain
rule.  At a nonstationary geometry the coordinate-curvature contribution is
\begin{equation}
 C_{ab}=\sum_i g_i^{(q)}\mathcal D_{iab},
 \label{eq:sparse-b-curvature}
\end{equation}
and symmetrized only after the contraction.  If \(\mathbf g^{(q)}\) and
\(\mathbf H^{(q)}\) are the internal-coordinate gradient and Hessian, the exact
Cartesian derivatives are
\begin{align}
 \mathbf g^{(x)} &= \Bmat^T\mathbf g^{(q)},\\
 \mathbf H^{(x)} &= \Bmat^T\mathbf H^{(q)}\Bmat+\mathbf C.
 \label{eq:hessian-chain-rule}
\end{align}
Conversely, on the full-row-rank active coordinate space,
\begin{equation}
 \mathbf H^{(q)}=(\Bmat^+)^T\bigl(\mathbf H^{(x)}-\mathbf C\bigr)\Bmat^+.
 \label{eq:hessian-back-transformation}
\end{equation}
Thus \(\Bmat'\) is not an auxiliary coordinate-selection device: it is the
geometric-curvature term required to transform Hessians at a nonstationary or
constrained geometry.  At an unconstrained stationary point
\(\mathbf g^{(q)}=0\), so \(\mathbf C=0\) and the familiar first-order
transformation is recovered.  The distinction matters for scans, constrained
optimizations, force-field construction away from equilibrium and any
trajectory on which Cartesian and internal Hessians are exchanged.  It does
not enlarge the SONIC contract: the consumer builds only the derivative slices
it needs and remains responsible for singular-chart handling and numerical
fallback policy.

\section{Validation Strategy}

The validation target is not merely that a coordinate list can be printed.  A
\SONIC\ coordinate definition must first pass construction-level checks.  The
final coordinate count and analytic \(\Bmat\) rank must match the vibrational
rank.  The current Python construction and a separate Fortran control
implementation must agree on coordinate families, final rank and
Wilson-\(\Bmat\) row space for the golden corpus.  Symmetry-adapted blocks must
close under the stored point-group operations, and the saved SALC coefficient
vectors must remain normalized.  These are internal implementation-level
checks, not a comparison with an external published coordinate generator.

The broader scientific motivation includes vibrational analysis, sampling and
semi-experimental refinement, but the validation layers must be separated.  The
present paper validates the coordinate contract itself and tests whether an
external Gaussian 16 calculation can consume \SONIC{}-generated coordinates.
The scope of the numerical evidence retained here is stated explicitly in
Sec.~\ref{sec:scope-limitations}.

\begin{table}[htbp]
\centering
\scriptsize
\setlength{\tabcolsep}{3pt}
\renewcommand{\arraystretch}{0.95}
\caption{Validation map separating scientific claims from implementation
diagnostics.}
\label{tab:validation-map}
\begin{tabular}{@{}p{0.23\linewidth}p{0.34\linewidth}p{0.33\linewidth}@{}}
\toprule
Evidence & What it tests & What it does not claim \\
\midrule
Internal corpus, Table~\ref{tab:validation} & Rank closure, analytic
\(\Bmat\)-row consistency, point-group block closure and agreement with a
separate Fortran control implementation. & External superiority over another
published coordinate generator. \\
\midrule
Local-pseudosymmetry corpus, Table~\ref{tab:local-salc-validation} & Local
\(A_1\)-first ordering, rigid-rotation invariance, substituted/fused-ring domain
records and independent Python/Fortran classification. & Replacement of the
exact molecular point group or a complete local-irrep decomposition. \\
\midrule
Gaussian-GIC motifs, Table~\ref{tab:frisch-si} & Automatic construction of
coordinate objects that otherwise appear as explicit GIC examples or dedicated
motifs. & A benchmark of Gaussian's internal optimizer. \\
\midrule
Gaussian \texttt{ReadAllGIC} optimizations, Table~\ref{tab:readallgic-opt} &
Whether a commercial Gaussian 16 parser and optimizer can consume the generated
coordinate block. & A universal iteration-count advantage for every molecule. \\
\midrule
Puckering and portable-export probes, Figs.~\ref{fig:puckering-scan} and
\ref{fig:cyclohexane-puckering-equivalence}, Table~\ref{tab:external-codes} & Reuse of
the same generated coordinate path as Gaussian GIC input or ordinary Cartesian
geometries for other electronic-structure codes. & A full DVR application. \\
\midrule
Construction-cost study, Table~\ref{tab:construction-cost} and
Fig.~\ref{fig:construction-scaling} & Empirical scaling of coordinate
construction and sparse-\(\Bmat\) evaluation on the present test corpus. &
Asymptotic proof for all molecular graph families. \\
\bottomrule
\end{tabular}
\end{table}

\begin{figure}[htbp]
\centering
\includegraphics[width=0.95\linewidth]{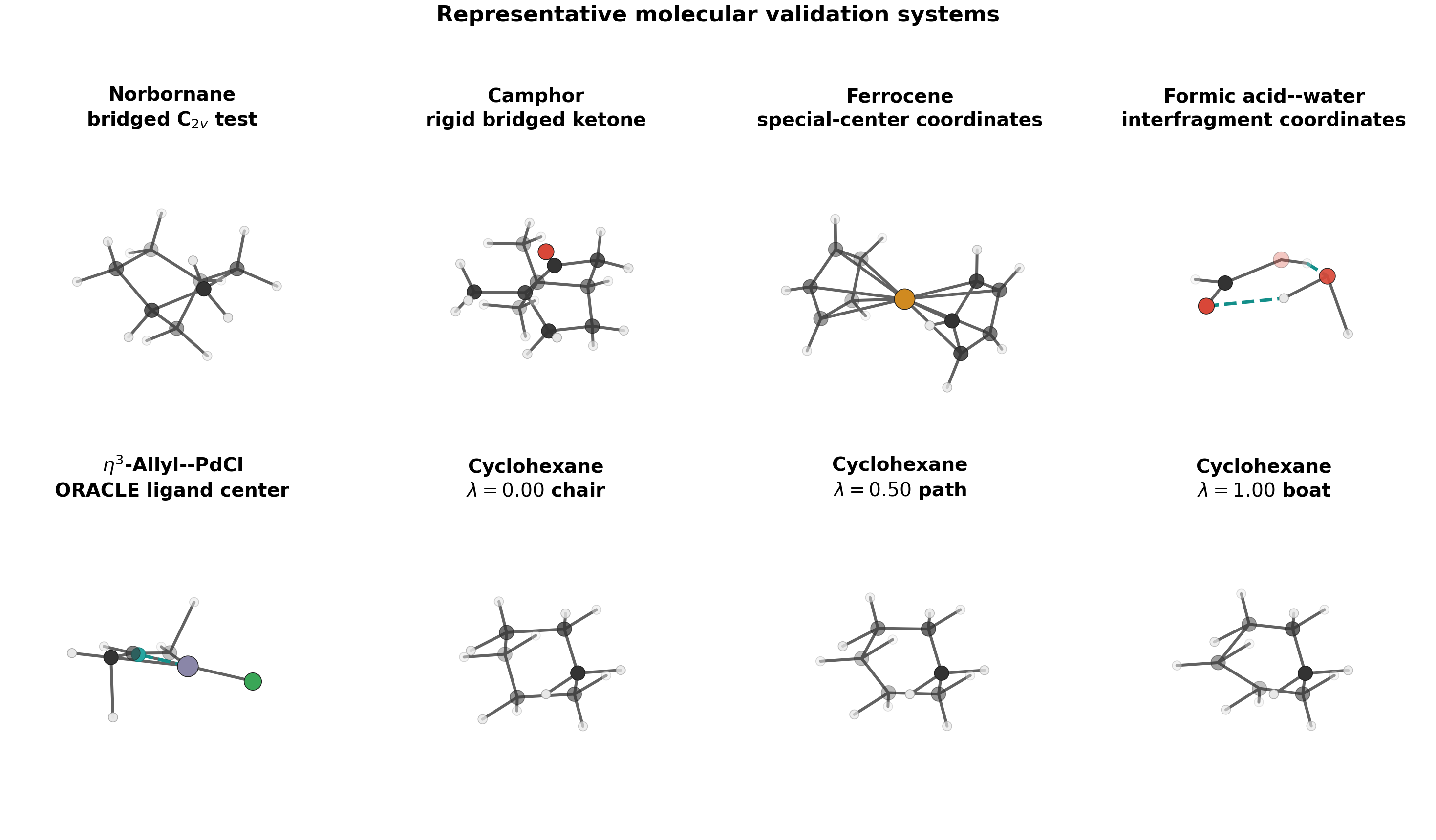}
\caption{Three-dimensional views of representative systems used in the
validation and application probes.
Norbornane and camphor test bridged-ring coordinate construction and
Gaussian-compatible improper-coordinate export; ferrocene tests center-based
special coordinates; formic acid--water tests relative fragment coordinates;
and $\eta^3$ allyl--PdCl tests a protected multi-atom ligand center.  Dashed
turquoise segments identify intermolecular contacts or a protected
center--metal coordinate, and the turquoise marker denotes the supplied
$\eta^3$ center.  The cyclohexane panels show three points along the chair-to-boat
puckering path used for the portable Cartesian/Gaussian \texttt{ReadAllGIC}
comparison.}
\label{fig:validation-molecules}
\end{figure}

Figure~\ref{fig:validation-molecules} shows representative molecular systems
used in the validation probes.  The figure is included to keep the coordinate
discussion tied to the underlying chemistry: bridged hydrocarbons test ring and
improper-coordinate choices, camphor adds a rigid asymmetric framework,
ferrocene exercises special center coordinates, the formic-acid--water complex
tests protected interfragment translations and orientations, the
$\eta^3$ allyl--palladium complex tests a supplied ligand center, and
the cyclohexane sequence tests portable puckering paths.

The regression corpus contains acyclic and cyclic molecules, fused and
non-benzenoid rings, bridged and spiro systems, high-symmetry cages, linear
molecules, ferrocene in \(D_{5h}\) and \(D_{5d}\) forms, a formic-acid--water
weak-complex probe, and an $\eta^3$ allyl--palladium interaction-center probe.  The
ring subset protects the \(RDef/RPck\) source spaces,
butterfly coordinates and the use of linear \(RPck\) rows rather than polar
active coordinates.  Ferrocene protects center-based special coordinates and
high-symmetry point-group projectors; the $\eta^3$ case protects consumption of an
explicit multi-atom ligand center.  Weak-complex tests protect the
fragment-coordinate policy and the analytic \(\Bmat\) rows of interfragment
translation and orientation coordinates.

Table~\ref{tab:validation} gives the self-consistency slice used for
this draft.  The first eight rows compare the present construction with a
separate Fortran control implementation; the reported residual is the
Wilson-row-space residual between two internally maintained non-redundant bases
that share the same scientific specification but not the same implementation
path.
The narrow range of
\(\Delta_\mathrm{PF}\) values is expected: these residuals are dominated by the
fixed row-space audit tolerance and floating-point floor of the comparison, not
by a molecule-specific physical error.  Their role is to show that the two
implementation paths saturate the same numerical equivalence criterion across
systems of different size and symmetry.  The formic-acid--water row uses the
\texttt{SPECIAL\_COORDINATES} fragment path, for which the covalent Fortran
control reference is not the relevant comparison; there the table reports direct
analytic \(\Bmat\)-rank closure and the number of protected special coordinates.
Because this weak-complex probe is \(C_1\), \(\epsilon_\mathrm{SALC}=0\) is
trivial by construction and should be read only as the absence of nontrivial
SALC blocks, not as a symmetry-adaptation test.  The semantic-layer row is
a unit-level contract test of the \texttt{PROTECT}/\texttt{OBSERVABLE} layer
introduced in the section on coordinate functions and downstream models; it
exercises duplicate rejection, row-equivalent substitution and the fail-fast
protected-prefix rule.

\begin{table}[htbp]
\centering
\scriptsize
\setlength{\tabcolsep}{3pt}
\caption{Representative internal validation of the present \SONIC\ construction.
Ranks are final/target ranks.  \(\Delta_\mathrm{PF}\) is the Python--Fortran
Wilson-row-space residual where the Fortran control comparison is applicable.
\(\epsilon_\mathrm{SALC}\) is the maximum norm error of nontrivial saved SALC
coefficient vectors, and \(\epsilon_\mathrm{FD}\) is the maximum componentwise
difference between a final analytic SONIC row and its central Cartesian finite
difference.}
\label{tab:validation}
\begin{tabular}{@{}p{0.17\linewidth}p{0.25\linewidth}p{0.07\linewidth}p{0.16\linewidth}p{0.27\linewidth}@{}}
\toprule
System & Stress test & PG & Rank/closure & Numerical check \\
\midrule
Benzene & Aromatic ring projector & \(D_{6h}\) & 30/30; 3 \(RPck\) & \(\Delta_\mathrm{PF}=6.67\times10^{-9}\); \(\epsilon_\mathrm{SALC}=2.22\times10^{-16}\) \\
Pyrrole & Heteroaromatic ring projector & \(C_{2v}\) & 24/24; 2 \(RPck\) & \(\Delta_\mathrm{PF}=6.92\times10^{-9}\); \(\epsilon_\mathrm{SALC}=2.22\times10^{-16}\) \\
Azulene & Non-benzenoid fused ring, butterfly & \(C_{2v}\) & 48/48; 6 \(RPck\) & \(\Delta_\mathrm{PF}=6.44\times10^{-9}\); \(\epsilon_\mathrm{SALC}=2.22\times10^{-16}\) \\
Norbornane & Bridged saturated ring & \(C_{2v}\) & 51/51; 3 \(RPck\) & \(\Delta_\mathrm{PF}=6.22\times10^{-9}\); \(\epsilon_\mathrm{SALC}=2.22\times10^{-16}\) \\
Cubane & High-symmetry cage & \(O_h\) & 42/42; cyclic bends & \(\Delta_\mathrm{PF}=6.41\times10^{-9}\); \(\epsilon_\mathrm{SALC}=2.22\times10^{-16}\) \\
Ferrocene & \(D_{5h}\) metallocene projector & \(D_{5h}\) & 57/57 & \(\Delta_\mathrm{PF}=6.71\times10^{-9}\); \(\epsilon_\mathrm{SALC}=3.33\times10^{-16}\) \\
Spiro system & Spiro ring source spaces & \(D_2\) & 87/87; 6 \(RPck\) & \(\Delta_\mathrm{PF}=6.97\times10^{-9}\); \(\epsilon_\mathrm{SALC}=2.22\times10^{-16}\) \\
Cyclooctane & Large-ring puckering & \(D_2\) & 66/66; 5 \(RPck\) & \(\Delta_\mathrm{PF}=9.84\times10^{-9}\); \(\epsilon_\mathrm{SALC}=2.22\times10^{-16}\) \\
Pagodane & Cage and fused-ring source spaces & \(D_{2h}\) & 114/114 & \(\epsilon_\mathrm{FD}=4.02\times10^{-10}\); stretches and cyclic bends covered \\
Indane & Fused aromatic/alicyclic rings & \(C_s\) & 51/51 & \(\epsilon_\mathrm{FD}=4.77\times10^{-10}\); butterfly and \(RPck\) rows covered \\
Formic acid--water & Weak-complex fragment model & \(C_1\) & 18/18; 6 protected & three translations plus three orientations; direct \(\Bmat\)-rank \(=18\) \\
Water dimer & Hydrogen-bonded fragment model & \(C_1\) & 12/12; 6 protected & standalone Cartesian input; three translations plus three orientations \\
Benzene--water & Aromatic--polar fragment model & \(C_1\) & 39/39; 6 protected & standalone Cartesian input; ring coordinates coexist with the fragment block \\
$\eta^3$ allyl--PdCl & Supplied $\eta^3$ ligand center & \(C_1\) & 24/24; 1 protected & \(q_{\eta^3}=1.8000\)~\AA; analytic center--atom row retained \\
Weak-complex pseudo-cycles & Intermolecular and intramolecular pseudo-bond
contacts & \(C_s\), \(C_{3v}\), \(C_1\) & rank closed & formic acid--water,
water dimer, Li\(_2\)--Ar, borane--ammonia and glycine I/II fixtures validate
the current pseudo-cycle/interfragment baseline \\
Semantic layer & \texttt{PROTECT}/\texttt{OBSERVABLE} contract & -- & unit probes & duplicate protected row rejected; row-equivalent substitution recorded; protected rank excess fails fast \\
\bottomrule
\end{tabular}
\end{table}

The finite-difference audit in the pagodane and indane rows is applied after
the complete stored SONIC chain rule, rather than separately to selected
primitive formulas.  It therefore covers local SALCs and the linear
combinations defining cyclic, butterfly and puckering coordinates.  For all 114
pagodane and all 51 indane coordinates, mathematical display vectors are
generated with the Wilson
matrix frozen at the reference structure,
\(\Delta\mathbf{x}_k=\mathbf{B}_0^+\mathbf{e}_k\Delta q_k\).  The two frames
\(\mathbf{x}_0\pm\Delta\mathbf{x}_k\) are therefore exactly symmetric and
freeze every other SONIC to first order; they are not presented as nonlinear
back-transformations.  The amplitude is reduced if necessary to preserve
perceived connectivity and proper orientation, keep the maximum atomic
displacement below 0.10~\AA, and change no reference bond by 15\% or more.
Both accepted frames are checked against the frozen input topology, and the
number of complete topology evaluations is retained in the diagnostic record.
An optional mass-weighted right inverse,
\(\mathbf{M}^{-1/2}(\mathbf{B}_0\mathbf{M}^{-1/2})^+\), provides the
corresponding minimum-mass-norm vector without changing this first-order
contract.  Its mass provenance is explicit: the user can select average atomic
masses, default spectroscopic isotopes, masses stored with a Cartesian Hessian,
or a named isotopologue; the automatic choice prefers Hessian masses when they
are available and otherwise uses average masses.  The mathematical step is
distinct from the interactive display
amplitude: the viewer normalizes each half-vector to a user-selected maximum
atomic displacement in \AA.  The machine record reports row-normalized
conditioning, Cartesian amplification, participation ratio, significantly
moving atoms and the dominant atom.  It also stores the center, ring or bond
domain, local pseudogroup, local irrep and independent local/global
total-symmetry flags whenever those labels apply.  Ring motions additionally
retain the
canonical ring order and primitive coefficients, which are highlighted with
their displacement arrows in the viewer.  The tests consume every generated
three-frame XYZ trajectory and vector table through the same data loader used
by the GUI.  The versioned \texttt{matrix.smith.sonic\_diagnostics.v2}
manifest has a distributed JSON Schema and dependency-free validator; the GUI
loader remains compatible with archived version-1 manifests.  For large
coordinate sets the same viewer can filter by coordinate family, local domain,
local pseudogroup and local or global total-symmetry character.

The local-pseudosymmetry validation is summarized separately in
Table~\ref{tab:local-salc-validation}.  The first probe is intentionally
globally \(C_1\): it contains three equivalent C--H bonds at one carbon and an
asymmetric C--F/C--O substituent framework.  The purpose is to verify that a
local \(C_{3v}\) methyl domain can be used to place its symmetric stretch and
bend combinations first without promoting the whole molecule to \(C_{3v}\).
A rigid rotation by 0.731 rad about the normalized axis
\((1,-2,0.5)\) leaves the complete center-domain diagnostic records identical.
The separate Fortran path independently reports the same \(C_{3v}\) center,
places \(A_1\) first and closes at 18 coordinates.  This test checks the local
classification contract; the established tetrahedral control test continues
to provide the stricter Python--Fortran ordered-primitive and Wilson-row check.
The high-coordination extension is tested on both template geometries for each
coordination number from five through nine.  Each geometry is repeated after a
small angular distortion of one ligand, giving 20 models in each implementation.
Python and Fortran retain the same local-group assignment and close at ranks
12, 15, 18, 21 and 24 for CN 5, 6, 7, 8 and 9, respectively.
For CN 10 and CN 12, for which no named template is assumed, both
implementations select the generic equivalence fallback and close at ranks 27
and 33.  Their Cartesian row-space residuals are $1.03\times10^{-8}$ and
$1.06\times10^{-8}$, respectively.  Rotation and ligand permutation leave
the accepted-template score and margin unchanged.  A separate near-tie stress
test, with a deliberately widened absolute acceptance gate, is classified as
\texttt{AMBIGUOUS} and therefore uses the same generic fallback.
Two chemically named controls complement the anonymous polyhedral probes: a
slightly distorted \(\mathrm{SF_6}\) structure gives the octahedral local block
and closes at 15/15, while idealized square-antiprismatic
\([\mathrm{ZrF_8}]^{4-}\) gives \(D_{4d}\) and closes at 21/21.  The former uses
explicitly relaxed descriptor and distance thresholds (0.1 and 0.05~\AA),
illustrating that pseudosymmetry is a declared construction choice rather than
an implicit relabelling.

\begin{table}[htbp]
\centering
\scriptsize
\setlength{\tabcolsep}{3pt}
\caption{Local-pseudosymmetry validation with the local-SALC path enabled.
The PG column is the exact molecular point group; the local column reports only
the center or ring domain relevant to the test.}
\label{tab:local-salc-validation}
\begin{tabular}{@{}p{0.19\linewidth}p{0.08\linewidth}p{0.22\linewidth}p{0.12\linewidth}p{0.29\linewidth}@{}}
\toprule
System & PG & Local domain & Rank & Diagnostic check \\
\midrule
\(C_1\) methyl probe & \(C_1\) & center \(C_{3v}\) & 18/18 & identical after rigid rotation; Python and Fortran put local \(A_1\) first \\
\midrule
Benzene & \(D_{6h}\) & ring \(D_6\) & 30/30 & 12 color-preserving cyclic operations \\
\midrule
Toluene & \(C_s\) & substituted ring \(C_1\) & 39/39 & one cyclic operation at the default descriptor/distance tolerances \\
\midrule
Azulene & \(C_{2v}\) & two fused-ring domains & 48/48 & ring domains 1 and 2 recorded; one shared edge assigned once \\
\midrule
CN 5--9 centers & \(C_1\) & ten local polyhedral templates & 12--24 & 20 ideal/distorted models; Python--Fortran group and rank agreement \\
\midrule
CN 10/12 centers & \(C_1\) & generic equivalence fallback & 27/33 & Python--Fortran row-space residual below \(1.1\times10^{-8}\) \\
\midrule
\(\mathrm{SF_6}\), \([\mathrm{ZrF_8}]^{4-}\) & \(C_1\), \(D_{4d}\) & octahedral and square-antiprismatic domains & 15/15, 21/21 & frozen template, score, margin and thresholds recorded \\
\bottomrule
\end{tabular}
\end{table}

Table~\ref{tab:frisch-si} gives the first validation slice aimed directly at
the recent Gaussian GIC examples.  These are not comparisons of optimizer
performance; they test whether the chemically meaningful variables illustrated
in the Supporting Information can be produced as \SONIC\ contract objects or
coordinate functions without hand-writing a molecule-specific GIC input block.
For the ferrocene row, the lower cyclopentadienyl ring is rotated through a
five-point torsional scan from the eclipsed \(D_{5h}\) form to the staggered
\(D_{5d}\) form.  The unprojected \SONIC\ contract closes the full rank at every
grid point, while the high-symmetry endpoints are covered separately by the
projector tests in Table~\ref{tab:validation}.  The sparse-\(\Bmat\) density
reported there is the fraction
\[
  n_{\mathrm{nz}}/(n_{\mathrm{GIC}}\,3N)
\]
of nonzero derivative entries in the stored Wilson matrix; it is a compactness
diagnostic for the sparse representation, not an additional rank criterion.

\begin{table}[htbp]
\centering
\scriptsize
\setlength{\tabcolsep}{3pt}
\caption{Targeted tests inspired by the Gaussian GIC Supporting Information.
The purpose is to test automatic \SONIC\ construction of the same kind of
coordinate object that Gaussian can evaluate when supplied as a GIC expression
or through a dedicated option.}
\label{tab:frisch-si}
\begin{tabular}{@{}p{0.20\linewidth}p{0.31\linewidth}p{0.41\linewidth}@{}}
\toprule
SI motif & \SONIC\ object & Check \\
\midrule
Formamide proton transfer & Generated \(R_{\mathrm{NH}}\), \(R_{\mathrm{OH}}\)
primitive signatures plus \(\Sigma r\) and \(\Delta r\) functions & Function
values and sparse chain-rule rows reproduce the sum and difference derivatives. \\
Li\(_2\)--Ar scan variables & Protected interfragment translation axes plus
polar radius and elliptic quadratic functions & \(R=\sqrt{x^2+y^2}\) and the
elliptic constraint propagate sparse derivatives from generated axes. \\
\(XY_3\) local block & Homogeneous \(C_{3v}\) bend block projected by the stored
point-group representation & The block gives one \(A_1\) and one degenerate
\(E\) pair without mixing with other coordinate families. \\
Weak-complex rotor variables & Protected fragment translations, fragment
orientations and fragment-center distances & The weak-complex model contains
automatic \(FTRANS\), \(FROT\) and \(FC\_DIST\) rows with nonzero sparse
orientation derivatives. \\
Ferrocene Cp torsion & Ring-center/Fe special coordinates retained through a
\(0,9,18,27,36^\circ\) Cp rotation scan & Each grid point has rank
\(57/57\), analytic \(\Bmat\) rank \(57\), stable ring-center identities and
sparse \(\Bmat\) density \(0.095\)--\(0.101\). \\
Cyclohexane ring flip coordinate & Automatically generated \(D_{3d}\)
ring-puckering block \(\mathrm{A}_{1g}RPck001+\mathrm{E}_uRPck001/002\) &
No hand-written \texttt{Gcyc}, \texttt{Ccyc} or \texttt{Scn0} input is supplied;
the model closes rank \(48/48\) and exports a Gaussian-readable puckering
model. \\
Reduced path subgroup & Same cyclohexane source state rebuilt with a requested
\(C_1\) working subgroup & The rank remains \(48/48\), while labels and
totally symmetric export are computed in the reduced group; this is the
intended path mode when a scan leaves the high-symmetry stratum. \\
\bottomrule
\end{tabular}
\end{table}

\section{Serialization and External-Consumption Probes}

The preceding tests establish rank closure and preservation of symmetry labels.
The compact application probes ask a different
question: whether an independent program can parse and use coordinates emitted
from the same frozen object.  They are deliberately small calculations because
their role is external consumption, not validation of an optimization or
potential-energy-surface algorithm belonging to SMITH.

\subsection{Gaussian 16 as an external GIC interpreter}

The most direct external test is to let Gaussian use the \SONIC\ coordinates as its
optimization coordinate system.  In this probe \SONIC\ generates the
Gaussian-readable \texttt{ReadAllGIC} block, and Gaussian reads that block,
evaluates the GIC derivatives with its own machinery and performs the
optimization.  Gaussian 16 is used specifically because its GIC language is
general enough to express sums, differences and nonlinear coordinate functions
and therefore tests SONIC definitions directly rather than through a bespoke
adapter.  No hand-written surrogate coordinates are introduced.  The three
systems in Table~\ref{tab:readallgic-opt} were chosen because they exercise
ring and cage coordinates while remaining small enough for repeated external
checks.  The native \SONIC\ contract can represent signed out-of-plane rows and
block-local multi-\(U\) \(RPck\) puckering rows.  Commercial Gaussian 16 has no native \(U\) out-of-plane
primitive, and the official Gaussian GIC documentation warns that active
generic coordinates combining multiple periodic dihedrals can behave poorly
because the periodicity of the component torsions is not applied inside the
composite expression.\cite{GaussianGICManual}  The commercial
\texttt{ReadAllGIC} export therefore uses the Gaussian 16 profile by default:
\(U\) rows are replaced by the corresponding
Gaussian improper dihedrals, composite \(RPck\) rows are not written to the
Gaussian input, and all terminally translated component \(D(\cdots)\) terms are added as active
redundant periodic coordinates.  After the Gaussian optimization, \SMITH\ reads
the optimized Cartesian structure and evaluates it again in the frozen
non-redundant \SONIC\ contract.  The calculations reported here used Gaussian
16 at HF/STO-3G.

The relevant observations are that Gaussian 16 accepts the automatically
generated coordinate block and reaches a stationary point in three to nine
steps for saturated ring and cage systems with different symmetry and rank.

\begin{table}[htbp]
\centering
\caption{Gaussian \texttt{ReadAllGIC} optimizations using \SONIC{}-generated
coordinate definitions.  The route was HF/STO-3G
\texttt{Opt=(ReadAllGIC,MaxCycle=80) NoSymm}.  Gaussian 16 does not print the
same \(NTRed/NRank\) diagnostic in these logs, so the Gaussian-rank column is
left blank.}
\label{tab:readallgic-opt}
\begin{tabular}{lccccc}
\toprule
System & PG & \SONIC\ rank & Gaussian \(NTRed/NRank\) & Steps & \(E_\mathrm{final}\) / \(E_h\) \\
\midrule
Cyclohexane & \(D_{3d}\) & 48/48 & -- & 5 & \(-231.482671171\) \\
Norbornane & \(C_{2v}\) & 51/51 & -- & 9 & \(-268.893823819\) \\
Cubane & \(O_h\) & 42/42 & -- & 3 & \(-303.781399977\) \\
\bottomrule
\end{tabular}
\end{table}

A separate comparison using \texttt{CalcFC}, reported below and reproduced in
the Supporting Information for norbornane, is distinct from this
no-\texttt{CalcFC} solver-consumption table.

A symmetry-enabled Gaussian 16 variant was run for norbornane without
\texttt{NoSymm}.  The
Gaussian input contains the full rank-complete \SONIC\ block required by
\texttt{ReadAllGIC}, but all non-totally symmetric rows are marked
\texttt{Frozen}.  Thus only the 15 \(A_1\) coordinates are active while the
remaining rows define the redundant-to-nonredundant transformation.  Gaussian
16 accepts this input and reaches a \(C_{2v}\)-preserving stationary point in
eight steps.

\begin{figure}[htbp]
\centering
\includegraphics[width=0.72\linewidth]{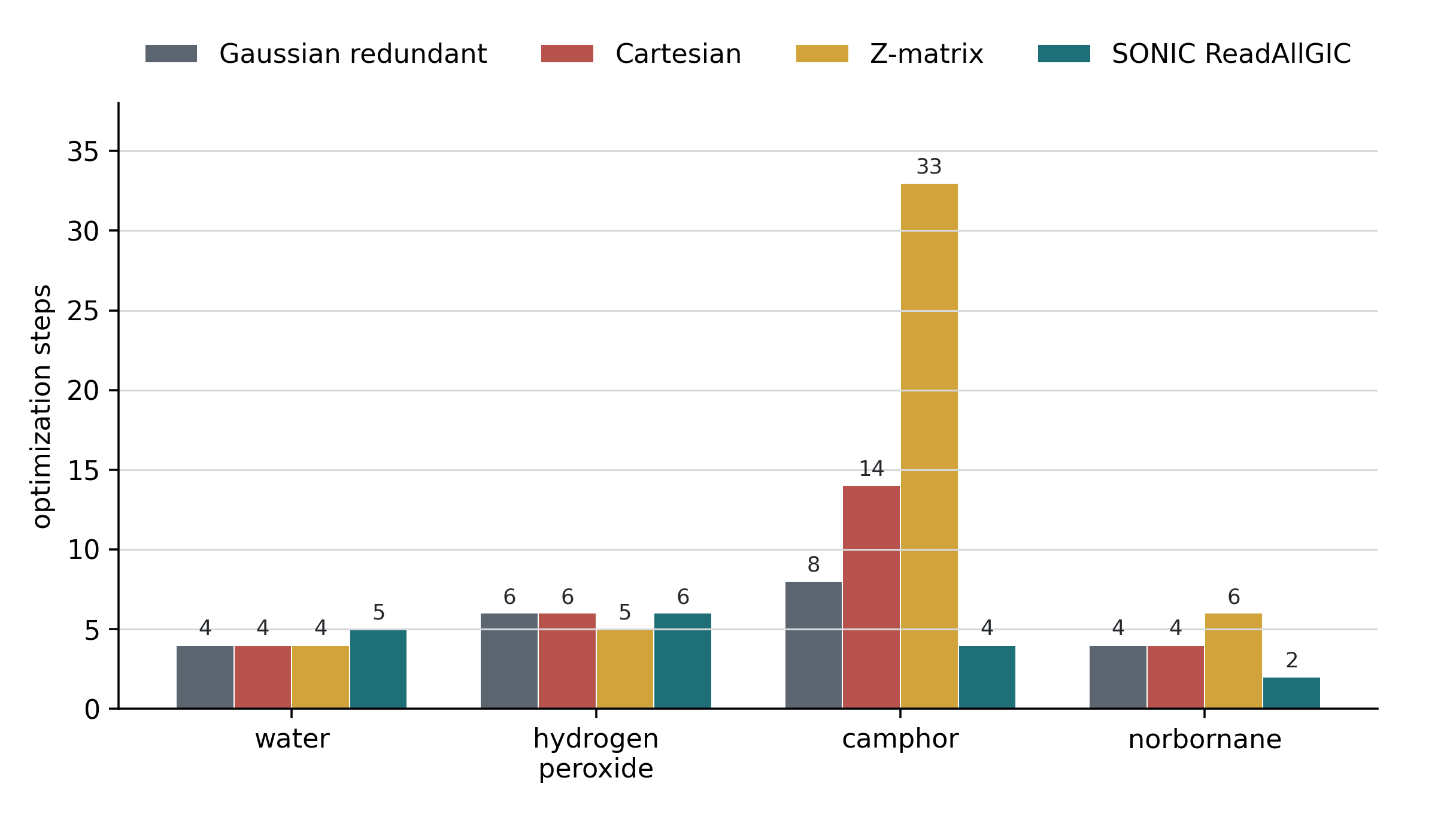}
\caption{Gaussian 16 HF/STO-3G optimization-step comparison for built-in
Gaussian coordinates and automatically generated \SONIC\ \texttt{ReadAllGIC}
coordinates.  The dataset shown here contains only completed runs; the plotting
script marks any future non-completed run with a cross rather than a zero-height
bar.  The water and hydrogen-peroxide tests verify solver consumption and input
portability; norbornane and camphor probe bridged-ring cases where a naive
sequential Z matrix is less efficient.}
\label{fig:gaussian-coordinate-comparison}
\end{figure}

The same export path was then compared directly with three standard Gaussian
optimization choices.  Two very small systems verify that no special handling
is needed when the coordinate problem is simple; norbornane and camphor add
larger bridged-ring cases, with camphor reaching 75 vibrational degrees of
freedom.  These calculations
are not a benchmark of electronic
structure theory or a claim that one coordinate choice always minimizes the
iteration count.  The \SONIC\ Gaussian input uses an initial force-constant
evaluation, as is common for difficult non-redundant coordinate optimizations.
They test the narrower point that \SMITH\ can generate a
Gaussian-readable \SONIC\ coordinate model without manual GIC transcription and
that Gaussian can optimize from it using the commercial \texttt{ReadAllGIC}
parser.  At HF/STO-3G, water and hydrogen peroxide give the same or nearly the
same iteration counts across the tested coordinate choices.  Norbornane
converges in four steps with Gaussian redundant internals and Cartesians, six
steps with a sequential Z matrix and two steps with the automatically
generated \SONIC\ block.  Camphor converges in eight steps with Gaussian
redundant internals, 14 Cartesian steps, four \SONIC\ compatibility steps and 33 steps for a
sequential Z matrix.  Table~\ref{tab:gaussian-coordinate-comparison} gives the
numerical values plotted in Fig.~\ref{fig:gaussian-coordinate-comparison}, so
the prose, table and figure refer to the same diagnostic set.  The point of this
diagnostic is therefore not that
\SONIC\ is always the shortest optimizer path in Gaussian, but that the
automatic generated GIC block is accepted by the commercial Gaussian
\texttt{ReadAllGIC} parser and remains competitive.  The commercial run is a
parser/optimizer compatibility probe, not a change in the scientific coordinate
contract: the non-redundant \SONIC\ rows remain the reference used to interpret
and compare output geometries.

\begin{table}[htbp]
\centering
\caption{Gaussian 16 HF/STO-3G optimization-step comparison shown in
Fig.~\ref{fig:gaussian-coordinate-comparison}.  The default column is Gaussian's
built-in redundant internal-coordinate optimizer.  The \SONIC{} column uses the
automatically generated Gaussian \texttt{ReadAllGIC} block.}
\label{tab:gaussian-coordinate-comparison}
\begin{tabular}{lccrrrr}
\toprule
System & PG & \SONIC\ rank & Default & Cartesian & Z matrix & \SONIC \\
\midrule
Water & \(C_{2v}\) & 3/3 & 4 & 4 & 4 & 5 \\
Hydrogen peroxide & \(D_{2h}\) & 6/6 & 6 & 6 & 5 & 6 \\
Norbornane & \(C_{2v}\) & 51/51 & 4 & 4 & 6 & 2 \\
Camphor & \(C_1\) & 75/75 & 8 & 14 & 33 & 4 \\
\bottomrule
\end{tabular}
\end{table}

A separate internal native-coordinate check was carried out with the \(U\)
out-of-plane representation retained.  This is not a commercial Gaussian 16
input: G16 requires its terminal translation to an improper dihedral.  For a three-coordinate center \(j\)
with ligands \(a,b,c\), the signed coordinate is
\begin{equation}
\label{eq:native-out-of-plane-u}
  U(j;a,b,c)=
  \arcsin\!\left[
  \frac{\hat r_{ja}\cdot(\hat r_{jb}\times\hat r_{jc})}
       {|\hat r_{jb}\times\hat r_{jc}|}
  \right],
\end{equation}
with the stored atom ordering fixing the sign convention.  In the native mode
camphor contains one native \(U(6,4,7,8)\) row and converges in eight
HF/STO-3G optimization steps with the internal test evaluator.  The commercial Gaussian 16 compatibility input
uses the Fortran/G16 improper-dihedral convention
\(U(I,J,K,L)\mapsto D(J,I,L,K)\), giving \(D(4,6,8,7)\) for the same local
non-planarity.  For ring puckering, commercial Gaussian 16 receives the
individual periodic dihedrals obtained by terminal translation of the
\(U\) sources of each \(RPck\) row as active redundant
coordinates; the original \(RPck\) combinations are retained only in the
\SMITH\ contract and do not appear in the commercial Gaussian 16 input.  The
optimized structure is then projected back onto the native \SONIC\ contract.
This compatibility export is less compact than the native
form, but it allows standard Gaussian 16 to be used without relying on
development-only primitives.  The native \SONIC\ contract still defines \(U\),
multi-\(U\) \(RPck\), \(QPck\) and \(\Phi P\).
For the symmetry-enabled norbornane export, the full rank-complete contract
contains 51 generated rows, of which 36 non-totally symmetric rows are marked
\texttt{Frozen}.  This directly tests frozen-irrep handling through the
commercial Gaussian 16 parser without requiring a nonstandard executable.

\subsection{Ring Puckering Branch}

This probe uses the same cyclobutane definition to define a one-dimensional
ring-puckering branch.  The coordinate is the alternating carbon-ring puckering
amplitude associated with the \(D_{2d}\) ring source block.  Single-point
Gaussian 16 HF/STO-3G energies were evaluated along a fixed substituent-frame
branch.  This is intentionally not a relaxed conformational profile; it is a
test that a named puckering coordinate can be used as a reproducible sampling
axis while the basis retains the same ring-source identity.

\begin{figure}[htbp]
\centering
\includegraphics[width=0.68\linewidth]{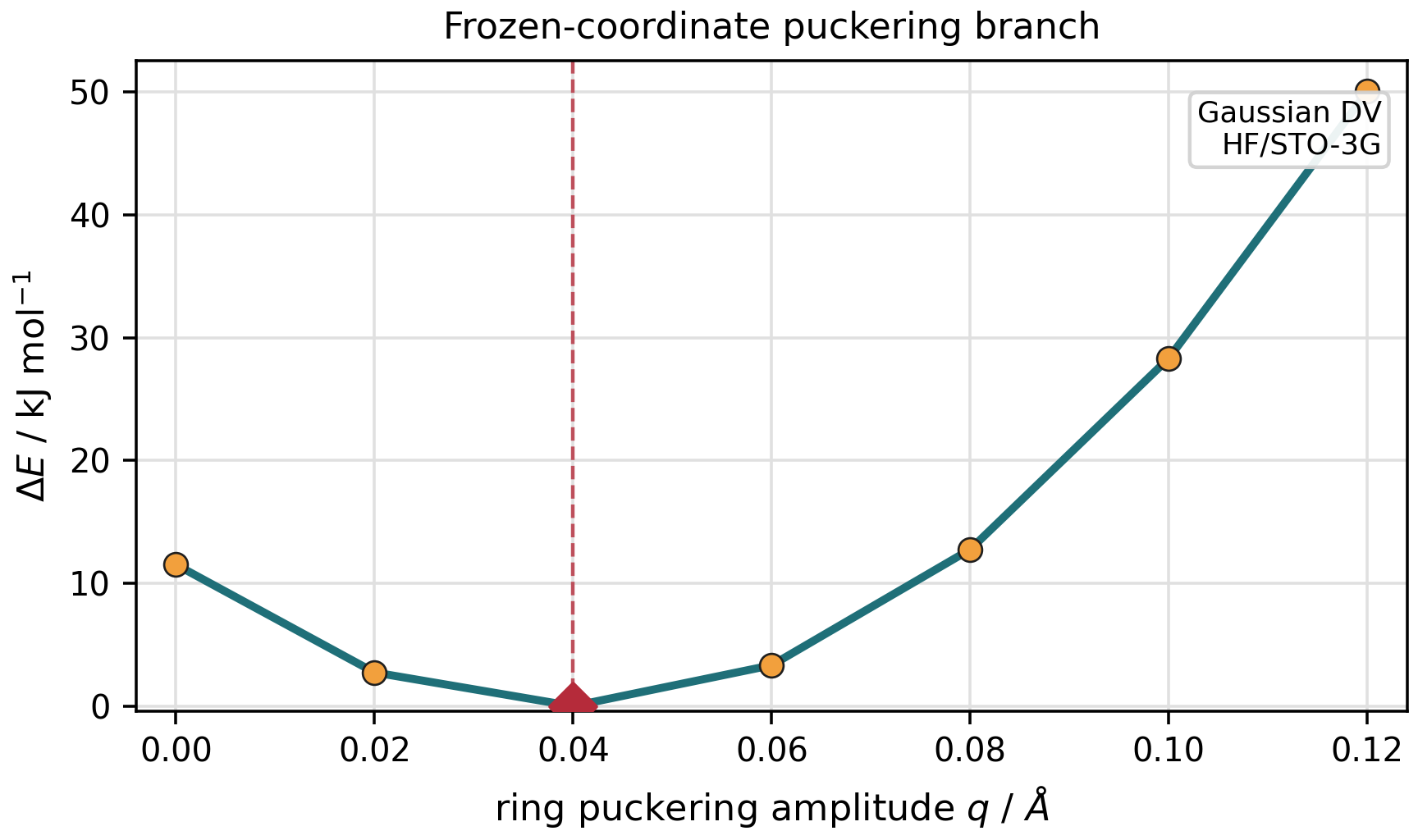}
\caption{Fixed-frame cyclobutane puckering branch generated from the frozen
\SONIC\ ring coordinate and evaluated with Gaussian 16 HF/STO-3G single-point
energies.  The branch is a coordinate diagnostic, not a relaxed conformational
energy profile.}
\label{fig:puckering-scan}
\end{figure}

Figure~\ref{fig:puckering-scan} illustrates the result.  The minimum on this
fixed branch is near \(q=0.04\)~\AA, matching the optimized reference
geometry.  Larger puckering displacements raise the energy smoothly.  In a DVR
or PES-fitting application, the same base coordinate could be replaced by a
Cremer--Pople amplitude or phase function \(y=f(\q)\), while the analytic
chain-rule row would still descend from the frozen \SONIC\ \(\Bmat\).

\subsection{Portable Chair-to-Boat Puckering Export}

The final probe tests a stricter export question, using the same cyclohexane
ring-flip motif discussed in the Gaussian GIC paper and its Supporting
Information.  There, the puckering coordinate is encoded by an explicit
molecule-specific input block containing variables such as \texttt{Gcyc},
\texttt{Ccyc} and \texttt{Scn0}.  Here no such block is provided.  \SONIC\ starts
from the molecular graph and geometry, generates the ring-puckering path,
and then constructs a fixed-frame cyclohexane puckering path from the chair
\(q_3\) component to the boat \(q_2\) component.  For each point it generated
two independent external representations of the same geometry: a portable
Cartesian input, usable by any electronic-structure program, and a Gaussian
\texttt{ReadAllGIC} input containing the automatically generated coordinate
contract.  Gaussian B3LYP/6-31+G* single-point energies were then evaluated
from both inputs.

\begin{figure}[htbp]
\centering
\includegraphics[width=0.78\linewidth]{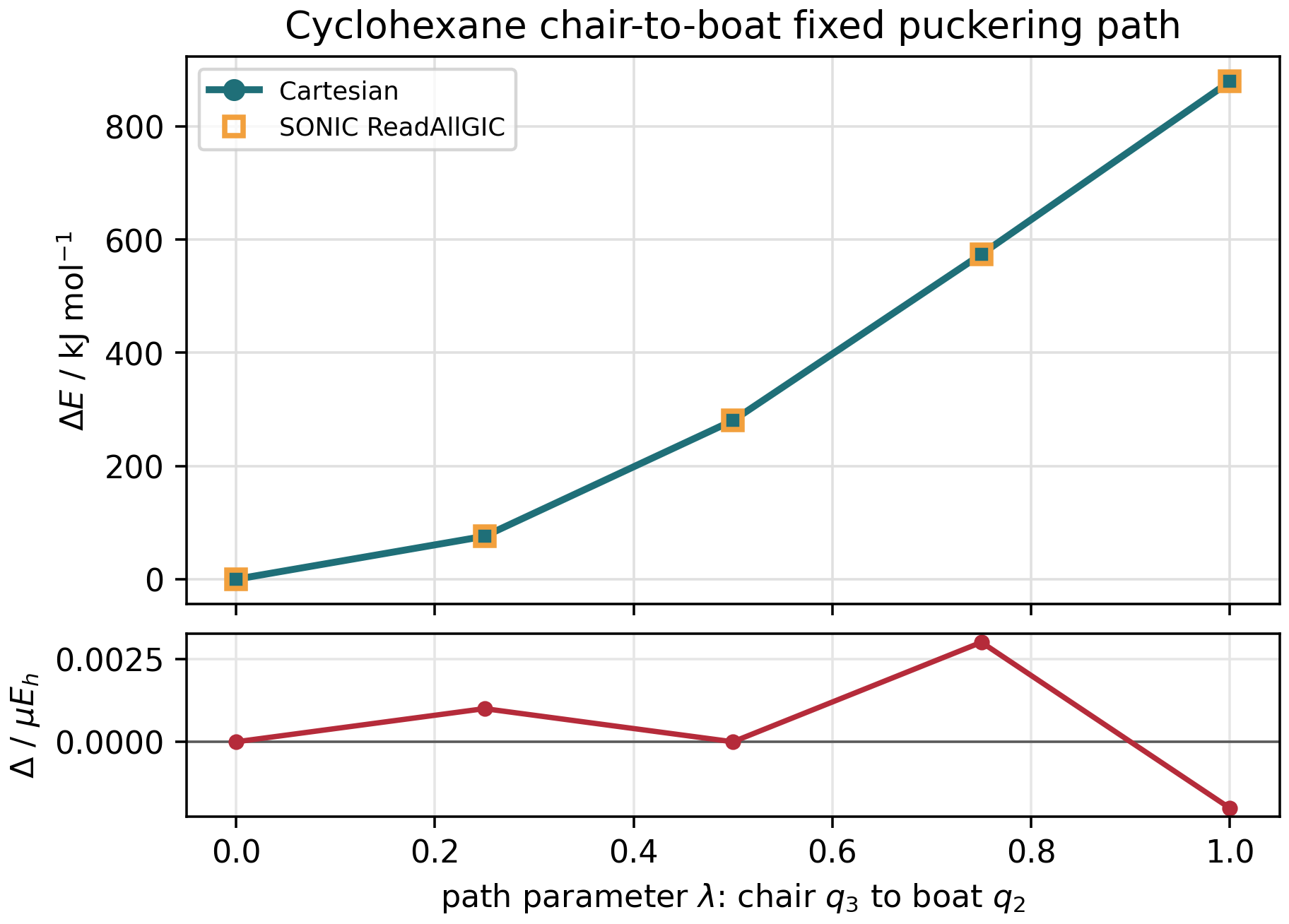}
\caption{Cyclohexane fixed-frame chair-to-boat puckering export test
(Gaussian B3LYP/6-31+G*).  The upper panel compares the same five geometries
evaluated from portable Cartesian inputs and from \SONIC{}-generated
\texttt{ReadAllGIC} inputs.  The lower panel gives the energy difference in
microhartree.}
\label{fig:cyclohexane-puckering-equivalence}
\end{figure}

All five points have rank \(48/48\) and contain no out-of-plane GIC terms, so
the comparison uses only syntax supported by standard Gaussian
\texttt{ReadAllGIC}.  Figure~\ref{fig:cyclohexane-puckering-equivalence} shows
that the two energy traces are numerically identical: the largest
Cartesian--\texttt{ReadAllGIC} difference is \(3.0\times10^{-3}\,\mu E_h\),
about \(8\times10^{-6}\)~kJ~mol\(^{-1}\).  The large rise along the path is not
a chair--boat barrier; it is the expected cost of an unrelaxed fixed-frame
puckering deformation.  The result instead shows that the same \SONIC{}-generated
large-amplitude coordinate path can be exported either as ordinary Cartesians or
as an automatically generated Gaussian GIC model without changing the electronic
energy surface being sampled.

Because the Cartesian branch is ordinary molecular geometry, it is not tied to
Gaussian.  The same three representative points were also evaluated on a local
workstation with ORCA and Molpro at PBE0/def2-SVP, a different level from
the B3LYP/6-31+G* Gaussian check.  The absolute energies differ slightly
between codes, as expected from independent DFT implementations and defaults,
but the relative fixed-frame puckering profile is the same at the level needed
for this export test (Table~\ref{tab:external-codes}).  This verifies the
intended division of labor: \SONIC\ constructs the model and coordinates, while
external electronic-structure programs consume the resulting Cartesian
geometries or GIC serializations.  It confirms that the coordinate definition is
independent of the electronic-structure implementation.

\begin{table}[htbp]
\centering
\caption{Portable cyclohexane puckering coordinates evaluated by non-Gaussian
codes on a local workstation.  Energies are PBE0/def2-SVP single points on the same
\SONIC{}-generated Cartesian geometries; relative energies are in kJ mol\(^{-1}\)
with respect to \(p00\).  The labels are the same interpolation points used in
Fig.~\ref{fig:cyclohexane-puckering-equivalence}: \(p00\) corresponds to
\(\lambda=0.00\), \(p02\) to \(\lambda=0.50\), and \(p04\) to
\(\lambda=1.00\).}
\label{tab:external-codes}
\begin{tabular}{lrrrr}
\toprule
Point & ORCA \(E_h\) & ORCA \(\Delta E\) & Molpro \(E_h\) & Molpro \(\Delta E\) \\
\midrule
\(p00\) & -235.422463982 & 0.000 & -235.422142275 & 0.000 \\
\(p02\) & -235.321345943 & 265.485 & -235.321038853 & 265.447 \\
\(p04\) & -235.103704314 & 836.903 & -235.103335256 & 837.028 \\
\bottomrule
\end{tabular}
\end{table}

\section{Discussion}

\SONIC\ can be viewed as a constructive generalization of natural internal
coordinates.  Its novelty is not the use of Wilson rows, SVD, Gram--Schmidt or
SALC projectors separately, but the deterministic protocol that combines them
while respecting locality, symmetry, semantic priority, protected coordinates,
ring source spaces and fragment coordinates.  The method is deliberately not a
single global orthogonalization: local and block-local decisions are followed by
a rank audit in the physical Wilson \(\Bmat\)-row space.  The structural
properties stated above are correspondingly modest: relabelling covariance,
determinism on a frozen state, protected-coordinate preservation when rank
allows it, block-local scaling invariance and local stability within a fixed
molecular-state stratum.  Each follows from the fixed candidate order,
normalized Gram--Schmidt rank tests and homogeneous point-group projectors.

The method should therefore be read as a documented coordinate-selection
protocol, not as a proof that only one internal-coordinate basis is possible.
Normalized Wilson-row testing removes the most direct dependence on raw physical
units in the final rank decision, but the protected-first family order remains a
chemical modelling choice.  The point is that this choice is explicit,
reproducible and reported, rather than hidden in primitive weights chosen before
a global diagonalization.

This is also the practical answer to the possible ``black box'' criticism: the
output stores primitive definitions, coefficients, family labels, irreducible
representations, rank diagnostics and analytic derivative rows.  Downstream
calculations can also compare the \GIC\ active space with a symmetry-adapted
Cartesian nullspace that is not built from GIC primitives; agreement shows that
the chemical conclusion depends on topology, symmetry and rank, not on a hidden
generator choice.

The second-order boundary prevents a different category error.  The ordinary
Wilson matrix describes the tangent map
\(d\mathbf q=\Bmat\,d\mathbf x\); its derivative describes how that tangent map
curves as the geometry changes.  Equations~\eqref{eq:hessian-chain-rule} and
\eqref{eq:hessian-back-transformation} explain why a Hessian-transforming
application may require \(\Bmat'\) away from equilibrium.  They do not make
\(\Bmat'\) a SONIC construction product.  At an exactly stationary harmonic
point the gradient contraction vanishes, while at a constrained or
nonstationary point the consuming program evaluates the required derivative
slices from the frozen coordinate formulas.  SMITH therefore remains a
first-derivative coordinate constructor rather than a general Hessian engine.
In MATRIX this second-order consumer is ARCHITECT.

Table~\ref{tab:local-salc-validation} adds a complementary check that is not
available from molecular point-group tests alone: a chemically useful local
totally symmetric coordinate remains identifiable in a \(C_1\) molecule, but
it does not acquire a false molecular \(A_1\) label.  The ring cases show the
same separation between a cyclic local subgroup and the exact molecular group.
The local thresholds can change a domain partition near a boundary; because
the tolerances, confidence, participating atoms and kept/pruned status are
serialized, that change is auditable rather than implicit.
The same table also removes the former high-coordination refactor boundary:
both implementations now recognize the two documented templates for every
coordination number from five to nine, while the generic construction closes
CN 10 and CN 12 without introducing additional named polyhedra.  The
ligand-equivalence fallback is also selected for unrecognized, ambiguous or
chemically split environments.  Because the score, second-template margin,
thresholds and selected status are serialized, this choice is fixed and
auditable across independent uses of the same coordinate contract.

The internal validation in Table~\ref{tab:validation} establishes rank closure,
symmetry-projector normalization and Python--Fortran row-space equivalence for
representative cases.  This is an implementation-level cross-check inside
the same development lineage, not a comparison with an external published
natural-coordinate program.  Table~\ref{tab:frisch-si} addresses the more
external-facing question raised by the Gaussian GIC Supporting Information: can
the same kinds of chemically specialized variables be generated without
hand-writing molecule-specific GIC expressions?  Gaussian supplies a powerful
expression and derivative language; \SONIC\ supplies automatic construction,
semantic interning of generated coordinates, sparse derivative propagation and
frozen provenance.  In this sense, GIC syntax is a serialization and solver
interface for \SONIC, not a competing coordinate-generation theory.  The
cyclohexane check makes this distinction concrete: the Gaussian-style puckering
model is available to the external solver, but its coordinate definitions are
not typed by hand.

\subsection{Computational Cost and Remaining Validation}

The expensive step is construction.  For \(m\) primitive candidates, target rank
\(r\), \(N\) atoms and \(z\) nonzero derivative entries, the current dense rank
audit costs \(O(mr\,3N)\) after local reductions have been formed.  In ordinary
molecular graphs \(m\) and \(r\) are \(O(N)\), giving a conservative cubic
build-time bound.  This is a pessimistic upper limit: ring, coordination and
fragment operations are confined to typed source blocks, and symmetry projection
is applied after rank reduction rather than to a full heterogeneous primitive
matrix.  Once the definition is stored, an application evaluates sparse Wilson
rows with cost \(O(z)\), plus any chain-rule work it elects to perform.  This
separation is the reason the code keeps coordinate construction and
\(\Bmat\) evaluation as distinct services.  Table~\ref{tab:construction-cost}
times only these SMITH operations; it contains no \(\Bmat'\), Hessian
transformation or internal-to-Cartesian realization benchmark.

\begin{figure}[htbp]
\centering
\includegraphics[width=\linewidth]{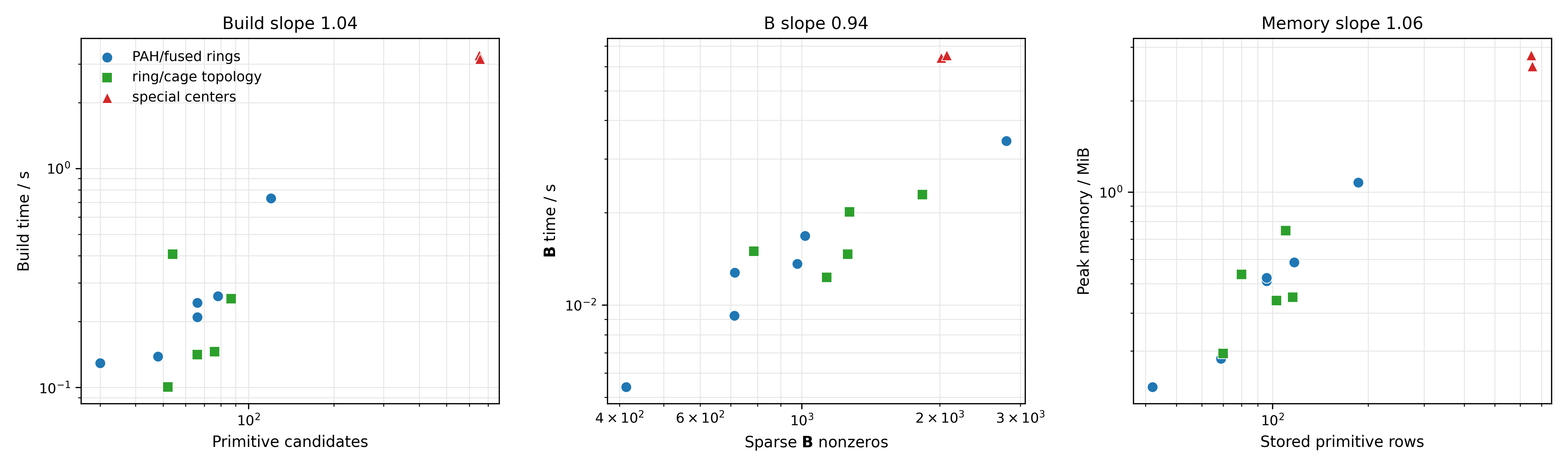}
\caption{Scaling diagnostics from Table~\ref{tab:construction-cost}.  The left
panel shows build time versus primitive-candidate count; the middle panel shows
one \(\Bmat\) evaluation versus sparse nonzero entries; the right panel shows
peak traced Python memory versus stored primitive rows.  The special-coordinate
ferrocene cases are separated because they deliberately keep a much larger
semantic source space than the final active rank.}
\label{fig:construction-scaling}
\end{figure}

Table~\ref{tab:construction-cost} and Figure~\ref{fig:construction-scaling}
give a first scaling study for the present implementation.  The values are
median wall times from five Python runs on a Linux workstation
(two AMD EPYC 7543 32-core processors, 128 logical
CPUs with simultaneous multithreading, and 1.0~TiB RAM), measured with
\texttt{tracemalloc}; they are therefore engineering diagnostics rather than
hardware-independent constants.  The timings were collected as single-process
Python construction jobs, not as parallel throughput benchmarks.  The test set
contains three slices: a controlled fused-PAH series from benzene to coronene,
a ring/cage topology set, and two metallocene special-coordinate cases.  The
intended split is visible.  Once the coordinate contract exists, one analytic
Wilson-\(\Bmat\) evaluation costs milliseconds to a few \(10^{-2}\)~s for the
tested covalent and ring systems.  Construction is larger because it includes
candidate generation, protected-first rank reduction, optional symmetry
projection and writing the frozen sections.

In the controlled PAH growth series, a log--log fit gives a build-time slope of
1.48 versus atom count (\(R^2=0.79\)) and a \(\Bmat\)-evaluation slope of 0.94
versus the number of sparse nonzero derivative entries (\(R^2=0.94\)).  The
left panel of Figure~\ref{fig:construction-scaling} reports a different
diagnostic fit: build time against primitive-candidate count over the
non-special systems.  That broader fit has slope close to unity but is
intentionally not used as a universal law, because cubane and the high-symmetry
PAHs show that topology and point-group projection can dominate a simple
one-variable model.  Across the non-special systems, peak Python memory scales
approximately linearly with the number of stored primitive rows (slope 1.06,
\(R^2=0.81\)).  The ferrocene entries are the expected expensive
cases in this set because the special-coordinate mode retains a large
ring-center/metal primitive source space through symmetry projection before
reducing to the 57-dimensional active contract.

\begin{table}[htbp]
\centering
\scriptsize
\setlength{\tabcolsep}{3pt}
\caption{Construction and \(\Bmat\)-evaluation scaling study for the present
\SMITH/\SONIC\ implementation.  Times are medians of five local Python runs.
The build column includes construction, rank reduction, optional symmetry
projection and section writing.  The \(\Bmat\) column is one analytic
Wilson-\(\Bmat\) evaluation from the stored contract.}
\label{tab:construction-cost}
\begin{tabular}{@{}llrrrrrr@{}}
\toprule
Series & System & \(N\) & Cand. & Rank & Build / s & \(\Bmat\) / s & Peak MiB \\
\midrule
PAH & Benzene & 12 & 30 & 30 & 0.131 & 0.0054 & 0.23 \\
PAH & Naphthalene & 18 & 48 & 48 & 0.139 & 0.0093 & 0.28 \\
PAH & Anthracene & 24 & 66 & 66 & 0.243 & 0.0139 & 0.51 \\
PAH & Phenanthrene & 24 & 66 & 66 & 0.211 & 0.0131 & 0.52 \\
PAH & Pyrene & 26 & 78 & 72 & 0.256 & 0.0162 & 0.59 \\
PAH & Coronene & 36 & 120 & 102 & 0.722 & 0.0343 & 1.08 \\
Ring/cage & Cubane & 16 & 54 & 42 & 0.400 & 0.0139 & 0.45 \\
Ring/cage & Norbornane & 19 & 52 & 51 & 0.099 & 0.0120 & 0.29 \\
Ring/cage & Cyclooctane & 24 & 66 & 66 & 0.142 & 0.0201 & 0.53 \\
Ring/cage & Camphor & 27 & 76 & 75 & 0.145 & 0.0151 & 0.44 \\
Ring/cage & Spiro & 31 & 87 & 87 & 0.253 & 0.0225 & 0.75 \\
Special & Ferrocene \(D_{5h}\) & 21 & 651 & 57 & 3.324 & 0.0642 & 2.83 \\
Special & Ferrocene \(D_{5d}\) & 21 & 656 & 57 & 3.182 & 0.0653 & 2.59 \\
\bottomrule
\end{tabular}
\end{table}

The study is still finite and implementation-specific.  It samples one PAH
growth series plus chemically diverse topology and special-coordinate probes;
peak memory is Python allocator memory observed during one build-plus-\(\Bmat\)
trial.  A release benchmark should add larger controlled series for fragment
count and for user-protected semantic coordinates, but the present data already
quantify the construction/use separation on the molecular classes exercised in
the paper.

Other validation items remain open.  The local-stability statement gives an
existence threshold \(\delta\) in terms of decision margins and Lipschitz
constants, but no general numerical values are reported for a real molecule.
The local-SALC probe now supplies one rigid-rotation invariant and fixed-threshold
classification test, but it is not yet a statistical tolerance-sensitivity
study over thermally distorted structures.
Near-threshold point-group perception has the same limitation: the
implementation records operation residuals, minimum margins and warning flags,
but quantitative tolerance-sensitivity tables remain future work.  Likewise,
the Gaussian tests are export and solver-consumption tests, not a comparison
with an external natural-coordinate generator.  The semantic-layer unit probes
cover protected-prefix rejection; a chemically realistic overconstrained
example would still be useful.

The main limitation is no longer the absence of pseudo-bonds in the code path,
but the current validated scope of the pseudo-cycle tests.  Protected fragment
and special-center coordinates are already suitable for weak complexes and
coordination motifs, while the pseudo-bond baseline covers typed hydrogen-bond
closures and bifurcated contacts.  What remains to be expanded is the corpus of
symmetry-rich intermolecular cycles and the numerical stress testing of those
pseudo-cycle rows under geometry changes.

\subsection{Scope and Limitations}
\label{sec:scope-limitations}

\SONIC\ is intended for high-accuracy quantum-chemical, spectroscopic and
refinement workflows in which the coordinate model carries chemical information
and must be reused across modules.  It is not designed as a large-scale
thousand-atom molecular-dynamics coordinate generator, a generic classical
force-field engine or a machine-learning featurizer.  Users who need a frozen,
auditable, symmetry-aware coordinate object for molecules, rings, fragments
or coordination motifs are the target audience; users who only need a
black-box optimizer coordinate for very large systems should normally use a
lighter optimizer-specific representation.

The construction is conditional on a fixed molecular-state stratum.  Bond
breaking or forming, proton transfer that changes the covalent graph, and
coordination rearrangements require separate contracts or a future
state-interpolation model.  An EVB-like or subspace-interpolation layer may be
possible, but it is not part of the present theory.

The coordinate definitions remain frozen, whereas their Wilson rows are
evaluated at the current geometry.  For fragment rotations, local-chart
rebasing also transports values and rows covariantly.  Large transformations
that change topology, fragment membership or symmetry stratum still require a
new contract; adaptive interpolation between such discrete molecular states is
outside the present scope.

The same locality qualification explains why \(\Bmat'\) is outside the SMITH
contract.  A second-order consumer can obtain it analytically from the stored
primitive formulas and SONIC coefficients, but that consumer must define its
own treatment of singular or unsupported charts and report any numerical
fallback.  Near linear-angle singularities, fragment-rotation chart boundaries
or a topology change, a divergent \(\Bmat'\) is coordinate curvature, not a
physical force constant and not a failure of SONIC construction.

Standalone operation has a related, explicit scope boundary.  Supplied
topology or primitive sections are authoritative and preserved, while the
bundled Cartesian frontend uses a revision-pinned ORACLE subset to construct
ordinary topology, point-group operations, atom permutations and PICs.  Results
that depend on advanced continuous atom equivalence, nondefault quasi-symmetry
decisions or special interaction centers require those data in the input state.
This is a capability distinction between input profiles, not a runtime
dependency of SMITH on a separate application.  It also means that changes to
the embedded perception subset require an explicit standalone version update
and renewed equivalence tests against ORACLE.

The priority-first rule is a chemically motivated modelling principle, not a
mathematical uniqueness result over all possible coordinate bases.  Alternative
priority orders would define different, still reproducible contracts.  The
claim made here is that the priority order is explicit, versioned and
diagnosed, and that special coordinates are not silently displaced by
ordinary coordinates when they are compatible with the target rank.

Finally, determinism is conditional on the frozen molecular state:
geometry, topology, symmetry operations, continuous atom perception, synthon
descriptors, tolerances, semantic grammar version and contract schema version.
A different perception model or tolerance policy may produce a different
contract; the scientific requirement is that such a change is visible in the
provenance rather than hidden in an implementation choice.

\section{Conclusions}

Starting from a frozen topology and redundant primitive/B contract, \SONIC\ provides a general, frozen and symmetry-aware construction of
non-redundant internal coordinates.  Its main contribution is a
reproducible coordinate object that integrates ingredients usually treated
separately: natural-coordinate locality, analytic Wilson-\(\Bmat\) rank control,
local \(A_1\)-first pseudosymmetry blocks, block-local point-group
symmetrization and special coordinates for
rings, fragments and interaction centers.  This is the coordinate object
that SMITH reports in human-readable form and can serialize for external use.
Vibrational analysis, optimization, sampling, force-field construction and
refinement are possible consumers, but their algorithms lie beyond this work.
Nonlinear coordinate functions such as inverse or exponential distances,
trigonometric angular terms and Cremer--Pople descriptors can be built on top of
this base by analytic chain-rule differentiation.

Molecular symmetry remains an input-state responsibility.  In MATRIX that
state is produced once by ORACLE and consumed unchanged by SMITH.  The
standalone package achieves independence by embedding a revision-pinned subset
of the same ORACLE topology, symmetry and ordinary-PIC kernel.  A plain XYZ can
therefore produce symmetry-adapted SONIC coordinates without a second
installation, while a supplied frozen state remains authoritative.  This
implementation duplication is deliberately versioned and tested so that it
cannot become an undocumented alternative definition of symmetry.

The present construction supports the intramolecular machinery needed for
acyclic, cyclic, fused, bridged, spiro and high-symmetry molecules, together
with special-coordinate and pseudo-bond models for weak complexes within the
validated baseline described above.  The next methodological step is to enlarge
the pseudo-cycle validation corpus for intermolecular systems, using the same
rank, symmetry and analytic-\(\Bmat\) invariants used by the rest of \SONIC.
The local-SALC validation additionally shows that center and ring
pseudosymmetry can order coordinates without changing the exact molecular
point group.  The Gaussian \texttt{ReadAllGIC} probes show that the present
contract is usable by an independent, general symbolic-coordinate evaluator.
Gaussian 16 is chosen for this direct test because its GIC syntax is unusually
general, not because SMITH depends on Gaussian or seeks to replace its internal
coordinate generator.
The puckering and portable chair-to-boat export probes illustrate why the same
stored definition is relevant to large-amplitude work within the scope defined
above.  Full application benchmarks remain outside the present construction
paper.  The cyclohexane export check further shows
that a \SONIC{}-generated puckering path can be represented either as ordinary
Cartesians or as Gaussian \texttt{ReadAllGIC} input with numerically identical
energies.  The frozen \SONIC\ contract should therefore be viewed not only as a
coordinate representation, but as a reusable molecular representation on which
future force-field, multilevel and potential-energy-surface methodologies can
naturally be built.

\section*{Reproducibility}

The calculations reported in the validation and application sections were
generated with standalone \SMITH\ and Python 3.13.11.  The immutable standalone
manuscript snapshot is tagged \texttt{v0.1.0rc6}.  Its dependency lock and
checksums are recorded with the archived submission.  Gaussian
optimization jobs used Gaussian 16; other Gaussian-family probes use the local
executable documented in the accompanying scripts.  The method and basis set
are stated with each probe.  The reported
construction tests use fixed inputs and no stochastic seeds.  The
Python implementation is checked against a separate Fortran control
implementation for the corpus rows where that comparison is applicable.  The
local-pseudosymmetry data can be regenerated with
\texttt{scripts/run\_local\_salc\_validation.py}; it records the descriptor and
distance tolerances, all final ranks, the rigid-rotation check and the
Fortran center-classification result.
The plain-XYZ water regression additionally verifies the embedded perception
boundary by requiring a $C_{2v}$ label and four serialized symmetry operations.

\section*{Data and Code Availability}

The manuscript repository at
\url{https://github.com/yogibubu/Smith} contains the scripts, JSON data files,
figures and Supporting Information used to generate the tables and application
plots, including \texttt{data/local\_salc\_validation.json} and its generating
script.  The same repository contains a reduced standalone Python package,
installation instructions and water, norbornane, formic-acid--water,
water-dimer, benzene--water and $\eta^3$ allyl--palladium examples.  Its
\texttt{smith-sonic build} command accepts a complete validated \texttt{xyzin}
state, an extended XYZ carrying \texttt{TOPOLOGY} or \texttt{PRIMITIVES}, or a
plain Cartesian input through the bundled perception frontend.  The provider-neutral
\texttt{--require-frozen-state} option requires the complete contract.  The
standalone distribution includes the revision-pinned topology, point-group
symmetry and ordinary-primitive frontend and
optional Gaussian \texttt{ReadAllGIC} export; installing or running another
application is not required.  The reproducible standalone environment is tagged
\texttt{v0.1.0rc6}. Repository revisions, the dependency lock and
\texttt{\#SMITH\_PROVENANCE} record the code and input profile.

\section*{Author Contributions}

Vincenzo Barone conceived the method, designed the \SONIC\ coordinate
contract, implemented the active code paths used here, performed the
calculations and wrote the manuscript.  External refinement and PES-fitting
consumers are discussed only as downstream uses of the coordinate definition.

\section*{Acknowledgments}

The author acknowledges the Gaussian GIC work that motivated the distinction
between coordinate-description languages and automatic generation, and the
ring-reconstruction work of Lema-Saavedra and Fern{\'a}ndez-Ramos that clarified
how puckering coordinates connect to closed molecular geometries.

\bibliography{references}

\end{document}